\documentclass[showpacs,floatfix,superscriptaddress]{revtex4}
\usepackage{graphicx}
\usepackage{epsfig}
\usepackage{bm}
\usepackage[utf8]{inputenc}
\usepackage{amssymb}
\usepackage{float}
\usepackage{amsmath}
\usepackage{dcolumn}
\usepackage{latexsym}
\usepackage{subfig}
\usepackage{amsthm}
\usepackage{array}
\usepackage{framed}
\usepackage{cancel}
\usepackage[colorlinks]{hyperref}
\usepackage[usenames,dvipsnames]{color}
\usepackage{amsmath}   % برای فرمول‌ها
\usepackage{booktabs}  % برای خطوط زیباتر در جدول

\hypersetup{
     breaklinks=true,
    pdfstartview={FitH},    % fits the width of the page to the window
    colorlinks=true,       % false: boxed links; true: colored links
    linkcolor=blue,          % color of internal links
    citecolor=red,        % color of links to bibliography
    filecolor=magenta,      % color of file links
    urlcolor=blue,           % color of external links
    anchorcolor=green,      % Color for anchor text
    linktocpage=true
}

\def\doi{http://doi.org}

\begin{document}

\title{Accretion Process as a Probe of Extra Dimensions in MOG Compact Object Spacetimes}

\author{Kourosh Nozari}
\email[]{knozari@umz.ac.ir (Corresponding Author)}
\author{Sara Saghafi}
\email[]{s.saghafi@umz.ac.ir}
\author{Zeynab Ramezanpasandi}
\email[]{z.ramezanpasandi@gmail.com}

\affiliation{Department of Theoretical Physics, Faculty of Science, University of Mazandaran,\\
P. O. Box 47416-95447, Babolsar, Iran}

\begin{abstract}
The idea of extra spatial dimensions arises from attempts to unify gravity with other fundamental interactions, develop a consistent theory of quantum gravity, and address open problems in particle physics and cosmology. Considerable attention has been devoted to understanding how such dimensions modify gravitational theories. One way to probe their impact is through the analytical study of astrophysical processes such as black hole accretion. Since accretion efficiently converts gravitational energy into radiation, this makes it a powerful tool to test modified gravity (MOG) theories and higher-dimensional frameworks via the behavior of dark compact objects like black holes, neutron stars, and white dwarfs. In this work, we investigate the dynamics of neutral particles around a higher-dimensional, regular, spherically symmetric MOG compact object, focusing on the innermost stable circular orbit (ISCO), energy flux, temperature, and differential luminosity. We further analyze the accretion of a perfect fluid onto the same object, deriving analytical expressions for the four-velocity and proper energy density of the inflowing matter. Our findings show that extra dimensions reduce the ISCO radius while enhancing the corresponding flux and temperature. Finally, by comparing the effective disk temperature $T_{\text{eff}}$ with Event Horizon Telescope (EHT) observations of Sgr A*, we argue that MOG and higher-dimensional corrections to the accretion disk properties could be close to the current threshold of detectability.
.
\vspace{12 pt}
\end{abstract}


\maketitle

\enlargethispage{\baselineskip}
\tableofcontents

\section{Introduction}\label{intro}

From a phenomenological point of view, dark compact objects, such as white dwarfs, neutron stars, and black holes, belong to a large class of astrophysical entities. Such objects could theoretically appear in models beyond the Standard Model of particle physics or in the context of extended gravity theories \cite{Cardoso:2019rvt}. The existence of these extreme objects is confirmed by recent observations, including the LIGO/Virgo detections of gravitational waves from binary black hole mergers \cite{LIGOScientific:2016aoc,LIGOScientific:2016sjg,LIGOScientific:2016dsl,LIGOScientific:2017bnn} and the Event Horizon Telescope (EHT) imaging of supermassive black holes at the centers of the Milky Way and the M87 galaxy \cite{EventHorizonTelescope:2019dse,EventHorizonTelescope:2019uob,EventHorizonTelescope:2019jan,EventHorizonTelescope:2019ths,EventHorizonTelescope:2019pgp,
EventHorizonTelescope:2019ggy,EventHorizonTelescope:2021bee,EventHorizonTelescope:2021srq}. Future developments in very-long-baseline interferometry (VLBI) and gravitational wave astronomy are therefore likely to reveal new kinds of compact objects.\\

Even though Albert Einstein's General Theory of Relativity (GR) has been remarkably successful in explaining a wide range of observations and forecasting extraordinary phenomena, it is still not considered a comprehensive explanation of gravitational interactions and associated cosmic events. Some of its shortcomings are highlighted by difficulties like re-producing the rotation curves of nearby galaxies \cite{Sofue:2000jx,Sofue2016}, figuring out the mass profiles of galaxy clusters \cite{Ettori2013,Voigt2006}, and dealing with the intrinsic singularities at black hole centers. Additionally, in order to explain the observed late-time accelerated expansion of the Universe, GR necessitates the introduction of a cosmological constant $\Lambda$ \cite{Weinberg1989,Peebles2003}. The Scalar-Tensor-Vector Gravity (STVG), also known as MOdified Gravity (MOG), was developed and proposed by John W. Moffat \cite{Moffat2006}. One promising path toward extending GR is to modify its geometric sector through alternative formulations. In this framework, three scalar fields are introduced: the mass of the vector field $\tilde{\mu}$, the effective gravitational coupling $G$, and the vector-field interaction strength $\xi$. Together, these three scalar fields determine the gravitational behavior of spacetime. Astrophysical observations have been successfully addressed by the MOG theory. For example, it does not require dark matter to reproduce galaxy cluster dynamics and rotation curves \cite{Brownstein2006a,Brownstein2006b,Brownstein2007,Moffat2013,Moffat2014,Moffat2015a}, and it is consistent with Planck 2018 cosmological data \cite{Moffat2021a}. Furthermore, a variety of solutions have been derived within the MOG framework, such as cosmological models \cite{Roshan2015,Jamali2018,Davari2021}, higher-dimensional extensions \cite{Cai2021}, rotating and static black hole spacetimes \cite{Moffat2015b}, and even time-dependent, inhomogeneous configurations of mass-energy \cite{Perez2019}. Extensive theoretical and observational studies have also been prompted by the theory to investigate its properties and implications in various contexts \cite{Moffat2021b,Guo2018,Moffat2020,Mureika2016,Saghafi2021,Saghafi:2021wzx,Perez:2017spz,Shukla:2022sti,Hussain:2015cga,Sharif:2017owq,Lee:2017fbq,Kolos:2020ykz,Hu:2022lek,John:2019was}. Ref. \cite{Moffat2021_} notably presents a class of solutions describing regular (i.e., nonsingular) rotating and static MOG dark compact objects, while Ref. \cite{Sau:2022afl} analyzes the corresponding shadow structures.\\

Particles are drawn toward a dark compact object through a process called accretion, which releases excess energy into the surrounding environment, resulting in a variety of astrophysical phenomena \cite{Kato:20008,Martnez:2014}, including quasars, intense radiation, and powerful jets. Rotating gas that gradually spirals inward toward a massive central body forms a flattened structure called an accretion disk, which usually forms around compact objects when interstellar matter is present; in fact, the disks are made up of gaseous material that moves in unstable bound orbits around the compact source \cite{Kato:20008,Martnez:2014}.When the right circumstances are met, the gas particles release gravitational energy, some of which takes the form of heat, as they descend into the compact object's gravitational potential. The inner part of the disk cools as a result of some of this heat being converted to radiation \cite{Kato:20008,Martnez:2014}. The emitted radiation spans across the electromagnetic spectrum and can be detected by radio, optical, and X-ray telescopes. Its properties are directly influenced by the dynamics of the gas particles, which themselves depend on the configuration and characteristics of the central mass. As a result, accretion disk emission spectra analysis offers important astrophysical information. Accretion disks of compact objects have gathered a lot of attention and been thoroughly examined in numerous studies because of their rich diagnostic potential \cite{Michel:1972,Jamil:2008bc,Babichev:2008dy,Mustafa:2023scj,JimenezMadrid:2005rk,Babichev:2008jb,Giddings:2008gr,Mustafa:2023ngp,Sharif:2011ih,John:2013bqa,
Abbas:2023vua,Debnath:2015yva,Ganguly:2014cqa,Mach:2013fsa,Mach:2013gia,Karkowski:2012vt,Rehman:2023hro,Nozari:2023enj,Yang:2015sfa,Babichev:2004yx,Babichev:2005py,Gao:2008jv,Barausse:2018vdb,Nozari:2020swx,Zheng:2019mem,Salahshoor:2018plr,Jiao:2016iwp}.\\

On the other hand, the study of gravity in models like braneworld models, which postulate the existence of extra dimensions, has attracted a lot of interest in recent decades. In these models, our well-known ($3+1$)-dimensional brane is embedded in a higher-dimensional spacetime of ($4+n$) dimensions, where $n$ is a compact spacelike dimension \cite{20}. One interesting aspect of braneworld theory is that it is possible that quantum gravity effects could be detectable in lab settings at TeV energy scales, and these models suggest that higher-dimensional black holes might be produced in high-energy experiments, like those at the Large Hadron Collider or through cosmic ray interactions. As higher-dimensional gravitational theories \cite{21} have been developed, studying black holes in such extended dimensional settings has become particularly interesting. Tangherlini \cite{22} was the first to extend the Schwarzschild black hole solution to higher dimensions. Later, Dadhich et al. \cite{23} obtained the earliest static, spherically symmetric black hole solution in the braneworld framework, which exhibits the same structure as the four-dimensional Reissner–Nordström black hole. The physics of black holes in higher dimensions proves to be far more diverse and intricate than in four dimensions \cite{24}. The problem of accretion onto TeV-scale black holes in higher dimensions was initially analyzed by Giddings and Mangano \cite{25} within a Newtonian approximation. Subsequently, Sharif and Abbas \cite{26} studied phantom energy accretion onto a five-dimensional charged black hole and demonstrated the validity of the cosmic censorship hypothesis. \\

Motivated by the above considerations, we turn our attention to the study of accretion disks around higher-dimensional regular MOG dark compact objects. In this context, we also analyze the dynamics of electrically neutral test particles in such spacetimes and investigate the associated energy flux and temperature distributions. The structure of the paper is as follows. In Section \ref{STVG}, we review the field equations of MOG and present the higher-dimensional regular MOG dark compact object along with its main properties. Section \ref{MTP} is devoted to the study of the motion of electrically neutral test particles moving in this spacetime and exploring the accretion disk around this object, including the analysis of the temperature profile, differential luminosity, and radiant energy flux. In Section \ref{ARMDCO} , we examine static spherically symmetric accretion. Finally, Section \ref{Conclu} provides our concluding remarks.

\section{Higher-Dimensional MOG Dark Compact Object}\label{STVG}

In the theory of STVG, the total action is defined as \cite{Moffat2006}
\begin{equation}\label{act}
S=S_{GR}+S_{M}+S_{\phi}+S_{S}\,,
\end{equation}
in which
\begin{equation}\label{sgr}
S_{GR}=\frac{1}{16\,\pi}\int d^{D}x\,\sqrt{-g}\,\frac{1}{G}\,R\,,
\end{equation}
\begin{equation}\label{sphi}
S_{\phi}=-\int d^{D}x\sqrt{-g}\left(\frac{1}{4}\,B^{\mu\nu}B_{\mu\nu}+V_{1}(\phi)\right)\xi\,,
\end{equation}
and
\begin{equation}\label{ss}
\begin{split}
S_{S} & =\int d^{D}x\,\sqrt{-g}\left[\frac{1}{G^{3}}\left(\frac{1}{2}\,g^{\mu\nu}\,\nabla_{_\mu}G\,\nabla_{_\nu}G-V_{2}(G)\right)
+\frac{1}{\tilde{\mu}^{2}G}\left(\frac{1}{2}\,g^{\mu\nu}\,\nabla_{_\mu}\tilde{\mu}\,\nabla_{_\nu}\tilde{\mu}-V_{3}(\tilde{\mu})\right)\right.\\
& \left.+\frac{1}{G}\left(\frac{1}{2}\,g^{\mu\nu}\,\nabla_{_\mu}\xi\,\nabla_{_\nu}\xi-V_{4}(\xi)\right)\right]\,.
\end{split}
\end{equation}
The indexes $M$, $\phi$, and $S$ stand for all possible matter sources, vector field $\phi_{\mu}$, and three scalar fields in the theory, respectively. $D$ refers to the dimensions of spacetime, $g_{\mu\nu}$ is the background metric tensor and $g$ is the corresponding determinant. $R$ is the Ricci scalar constructed by contracting $R_{\mu\nu}$ as the Ricci tensor, $\xi$ is the vector field coupling, $V_{1}(\phi)$, $V_{2}(G)$, $V_{3}(\tilde{\mu})$, and $V_{4}(\xi)$ are the corresponding potentials of the vector field $\phi^{\mu}$ and three scalar field $G$, $\tilde{\mu}$, and $\xi$, respectively and $B_{\mu\nu}=\partial_{\mu}\phi_{\nu}-\partial_{\nu}\phi_{\mu}$. Also $\nabla_{\mu}$ stands for the covariant derivative in the background spacetime.

One can find the full field equations of the STVG framework by variation of the action $S$ concerning the inverse of the metric tensor, which yields \cite{Moffat2006}
\begin{equation}\label{G}
G_{\mu\nu}+G\left(\nabla^{\gamma}\,\nabla_{\gamma}\frac{1}{G}\,g_{\mu\nu}-\nabla_{\mu}\nabla_{\nu}\frac{1}{G}\right)=8 \pi G T_{\mu\nu}\,,
\end{equation}
in which $G_{\mu\nu}$ is the Einstein tensor defied as $G_{\mu\nu}=R_{\mu\nu}-\frac{1}{2}\,g_{\mu\nu}R$ and we have set $c=1$.

In the STVG theory, the total stress-energy tensor is defined as
\begin{equation}\label{sm}
T_{\mu\nu}={}^{(M)}T_{\mu\nu}+{}^{(\phi)}T_{\mu\nu}+{}^{(S)}T_{\mu\nu}
\end{equation}
in which ${}^{(M)}T_{\mu\nu}$ is the stress-energy tensor of matter sources, ${}^{(S)}T_{\mu\nu}$ is the stress-energy tensor of the scalar fields, and the stress-energy tensor of the vector field is
\begin{equation}\label{setphi}
{}^{(\phi)}T_{\mu\nu}=-\frac{1}{4}\left(B_{\mu}^{\,\,\,\sigma}B_{\nu\sigma}-\frac{1}{4}\,g_{\mu\nu}B^{\sigma\lambda}B_{\sigma\lambda}\right)\,,
\end{equation}
for which $V_{1}(\phi)=0$.

As has been assumed in \cite{Moffat2021_}, for a regular, static, and spherically symmetric higher-dimensional MOG dark compact object the vector field is massless, i.e., $\tilde{\mu} = 0$, the vector field coupling is taken to be unity i.e., $\xi = 1$, the gravitational source charge of the vector field is $Q_{g}=\sqrt{\alpha G_{N}}M$, where $M$ represents the gravitational source mass and $G_N$ is the Newton’s gravitational constant. The gravitational coupling is defined as $G=G_{N}(1+\alpha)$, in which the STVG parameter $\alpha$ refers to the deviation from Newtonian gravity. In this paper, we set $G_{N}$ = 1. Based on these assumptions, it can be concluded that $S_{M}=S_{S}=0$, and thus ${}^{(M)}T_{\mu\nu}={}^{(S)}T_{\mu\nu}=0$. Therefore Eqs. \eqref{G} and \eqref{sm} reduce to the following form respectively
\begin{equation}\label{rG}
G_{\mu\nu}=8\pi(1+\alpha){}^{(\phi)}T_{\mu\nu}\,,
\end{equation}
\begin{equation}\label{rT}
T_{\mu\nu}={}^{(\phi)}T_{\mu\nu}\,.
\end{equation}

As presented in \cite{Nozari2024}, the line element of a static, spherically symmetric higher-dimensional MOG dark compact object, with the metric signature $(+,-,-,-)$, is given by
\begin{equation}\label{ds}
ds^{2}=f(r)dt^{2}-\frac{1}{f(r)}dr^{2}-r^{2}d\Omega_{D-2}^{2}\,,
\end{equation}
where \( d\Omega_{D-2}^2 = d\theta_1^2 + \sin^2\theta_1\, d\theta_2^2 + \ldots + \prod_{i=1}^{D-3} \sin^2\theta_i\, d\theta_{D-2}^2 \)
 is the line element of the \( (D - 2) \)-dimensional unit sphere and we will denote $\theta_{D-2}$ by $\varphi$. $f(r)$ is defined as \cite{Nozari2024}
\begin{equation}\label{f}
f(r) = 1
- \frac{(D - 2) m r^{2(D - 3)} \omega_{_{D-2}}}{8\pi \left( r^{2(D - 3)} + \frac{(D - 2)^2 m^2 \alpha (1+\alpha) \omega_{_{D-2}}^2}{256 G^2 \pi^2} \right)^{3/2}}
+ \frac{(D - 3)(D - 2) G q^2 r^{2(D - 3)} \omega_{_{D-2}}^2}{32\pi^2 \left( r^{2(D - 3)} + \frac{(D - 2)^2 m^2 \alpha (1+\alpha) \omega_{_{D-2}}^2}{256 G^2 \pi^2} \right)^2}\,,
\end{equation}
where \emph{m} and \emph{q} are defined by
\begin{equation}
m \equiv \frac{16 \pi G M}{(D - 2) \omega_{_{D-2}}}, \quad
q \equiv \frac{8 \pi Q}{\sqrt{2(D - 2)(D - 3)}\, \omega_{_{D-2}}},
\end{equation}
and \(\omega_{_{D-2}} = \frac{2 \pi^{\frac{D-1}{2}}}{\Gamma\left(\frac{D-1}{2}\right)}\) refers the volume of the unit \((D - 2)\)-sphere.\\
It is clear that the metric function in Eq.~(\ref{f}) reduces to the MOG dark compact object solution obtained by Moffat~\cite{Moffat2021_} when restricted to $D=4$. In this framework, the MOG dark compact object admits a critical parameter $\alpha_{crit}=0.674$  \cite{Moffat2021_}, such that for $\alpha \leq \alpha_{crit}$ the geometry exhibits two horizons. An important point is that the vector field associated with the spin-$1$ graviton generates a repulsive gravitational interaction, which prevents the dark compact object in MOG from collapsing into a horizon-forming MOG black hole.

By setting $\alpha=0$ in the line element \eqref{ds}, one recovers the Schwarzschild–Tangherlini black hole solution in GR. Furthermore, the asymptotic behavior of the higher-dimensional MOG compact object in the limit $r\rightarrow\infty$ can be expressed as
\begin{equation}\label{fleabl}
f(r)\approx 1-\frac{2(1+\alpha)m}{r^{D-3}}+\frac{\alpha(1+\alpha)Gq^{2}}{r^{2(D-3)}},.
\end{equation}
For $\alpha \leq \alpha_{crit}$, the regular higher-dimensional static and spherically symmetric MOG dark compact object admits two horizons in the asymptotic region, given by
\begin{equation}\label{hrz}
r_{\pm}=\Big(M+M\alpha \pm \sqrt{M^{2}+M^{2}\alpha}\Big)^{\frac{1}{D-3}},,
\end{equation}
where $r_{-}$ denotes the inner (Cauchy) horizon and $r_{+}$ corresponds to the outer (event) horizon. In the special case $\alpha=0$, these horizons coincide and reproduce the event horizon of the Schwarzschild–Tangherlini black hole~\cite{Tangherlini:1963}. For $\alpha>\alpha_{crit}$, however, the solution describes a horizonless, regular, spherically symmetric MOG dark compact object.

\section{Motion of a test particle in a higher-dimensional MOG dark compact object spacetime}\label{MTP}
In this section, we investigate the equations of motion through Lagrangian formalism. Under temporal translation and rotation around the axes of symmetry, the line element \eqref{ds} is invariant. Therefore one can obtain two Killing vectors for spacetime of the higher-dimensional MOG dark compact object as follows \cite{Nozari2024}
\begin{equation}\label{killvec}
\begin{split}
& {}^{(t)}\zeta^{\mu}\frac{\partial}{\partial x^{\mu}}=(1,0,0,0)\frac{\partial}{\partial x^{\mu}}=\frac{\partial}{\partial t}\,,\\
& {}^{(\varphi)}\zeta^{\mu}\frac{\partial}{\partial x^{\mu}}=(0,0,0,1)\frac{\partial}{\partial x^{\mu}}=\frac{\partial}{\partial\varphi}\,,\\
\end{split}
\end{equation}
where each of them corresponds to a constant quantity (the specific energy and the specific angular momentum
respectively) for the motion of the test particle in this spacetime.

\subsection{Effective potential}
The Lagrangian of a test particle moving in the spacetime of the regular higher-dimensional MOG dark compact object is written as
\begin{equation}\label{lag}
\mathcal{L}=\frac{1}{2}g_{\mu\nu}\dot{x}^{\mu}\dot{x}^{\nu}\,,
\end{equation}
where over-dot stands for derivative with respect to the affine parameter $\tau$ and $\dot{x}^{\mu}\equiv u^{\mu}=(u^{t},u^{r},u^{\theta},u^{\varphi})$ is the four-velocity of the test particle. Since we want to study the thin accretion disk, we consider the planar motion of the particle on the equatorial plane with $\theta_{i}=\frac{\pi}{2}$. By writing the Euler-Lagrange equation
\begin{equation}\label{eullag}
\frac{d}{d\tau}\left(\frac{\partial\mathcal{L}}{\partial\dot{x}^{\mu}}\right)-\frac{\partial\mathcal{L}}{\partial x^{\mu}}=0\,,
\end{equation}
the components of the four-velocity as a function of constants of motion are obtained as follows
\begin{equation}\label{utE}
u^{t}=\frac{dt}{d\tau}=\dot{t}=\frac{E}{f(r)}=\frac{E}{1
- \frac{(D - 2) m r^{2(D - 3)} \omega_{_{D-2}}}{8\pi \left( r^{2(D - 3)} + \frac{(D - 2)^2 m^2 \alpha (1+\alpha) \omega_{_{D-2}}^2}{256 G^2 \pi^2} \right)^{3/2}}
+ \frac{(D - 3)(D - 2) G q^2 r^{2(D - 3)} \omega_{_{D-2}}^2}{32\pi^2 \left( r^{2(D - 3)} + \frac{(D - 2)^2 m^2 \alpha (1+\alpha) \omega_{_{D-2}}^2}{256 G^2 \pi^2} \right)^2}}\,,
\end{equation}
\begin{equation}\label{uphi}
u^{\varphi}=\frac{d\varphi}{d\tau}=\dot{\varphi}=\frac{L}{r^{2}}\,,
\end{equation}
where $E$ and $L$ are the spacific energy and the specific angular momentum per unit mass of the particle, respectively. using the Euler-Lagrange equation, we obtain
\begin{equation}\label{ur}
\begin{split}
v\equiv u^{r}&=\frac{dr}{d\tau}=\dot{r}=\left[-f(r)\left(1-\frac{E^{2}}{f(r)}+\frac{L^{2}}{r^{2}}\right)\right]^{\frac{1}{2}}\,,
\end{split}
\end{equation}
where $v$ denotes the radial component of the four-velocity. Utilizing the normalization condition $u^{\mu}u_{\mu}=1$ and Eqs. \eqref{utE} and \eqref{ur}, we can find the equation of motion for a massive particle in the following form
\begin{equation}\label{rdot}
\dot{r}^{2}=E^{2}-V_{eff}\,,
\end{equation}
in which $V_{eff}$ is the effective potential of the test particle described by the metric function and angular momentum as
\begin{equation}\label{veff}
\begin{split}
V_{\text{eff}} &= f(r) \left(1 + \frac{L^2}{r^2} \right) = \left( 1 - \frac{(D - 2)m r^{2(D-3)} \omega_{_{D-2}}}{8 \pi \left(r^{2(D-3)} + \frac{(D-2)^2 m^2 \alpha (1+\alpha) \omega_{_{D-2}}^2}{256 G^2 \pi^2}\right)^{3/2}} \right. \\&\quad + \left. \frac{(D - 3)(D - 2)G q^2 r^{2(D-3)} \omega_{_{D-2}}^2}{32 \pi^2 \left(r^{2(D-3)} + \frac{(D-2)^2 m^2 \alpha (1+\alpha) \omega_{_{D-2}}^2}{256 G^2 \pi^2}\right)^2} \right)
\left( 1 + \frac{L^2}{r^2} \right).
\end{split}
\end{equation}

\subsection{Stable circular orbits}

To describe the motion of matter in accretion disks, it is very useful to analyze the circular orbits around the central object. The properties of these orbits, such as energy and stability, directly affect the structure of the accretion disk and the emitted radiation. The condition  $\dot{r}=\ddot{r}=0$, which yields $\frac{dV_{eff}}{dr}=0$, determines the radius of circular orbits in a specific spacetime. Based on this condition and using Eqs. \eqref{utE} and \eqref{uphi}, we obtain the following relations for the specific energy ($E$), specific angular momentum ($L$), and the angular velocity ($\Omega_{\varphi}\equiv\frac{d\varphi}{dt}=\frac{u^{\varphi}}{u^{t}}$) for a test particle in the circular orbits in higher-dimensional MOG dark compact object background
\begin{equation}\label{E2}
\begin{split}
E^{2} & =\frac{2f^{2}(r)}{2f(r)-rf'(r)}\,,
\end{split}
\end{equation}
\begin{equation}\label{L2}
L^{2}=\frac{r^{3}f'(r)}{2f(r)-rf'(r)}\,,
\end{equation}
\begin{equation}\label{Omega2}
\Omega_{\varphi}^{2}=\frac{1}{2r}f'(r)\,,
\end{equation}
where a prime stands for differentiation with respect to the radial coordinate $r$. Because of the complicated form of the resulting equations, for the sake of economy the relations of $f(r)$ and $f'(r)$ are not applied. This choice is followed consistently until the end.

\begin{figure}[htb]
\centering
\subfloat[\small{\emph{}}]{\includegraphics[width=1\textwidth]{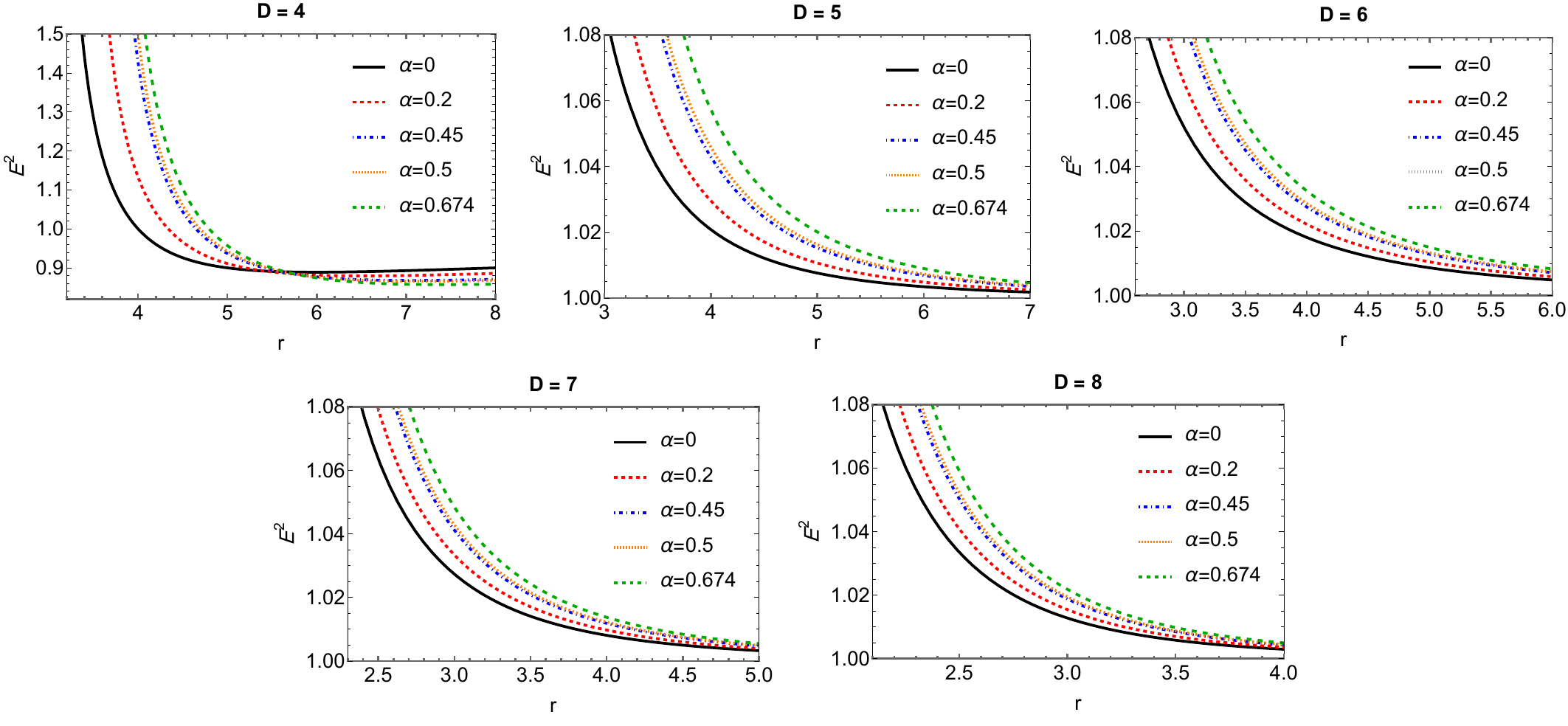}
\label{figE-D}}

\subfloat[\small{\emph{}}]{\includegraphics[width=1\textwidth]{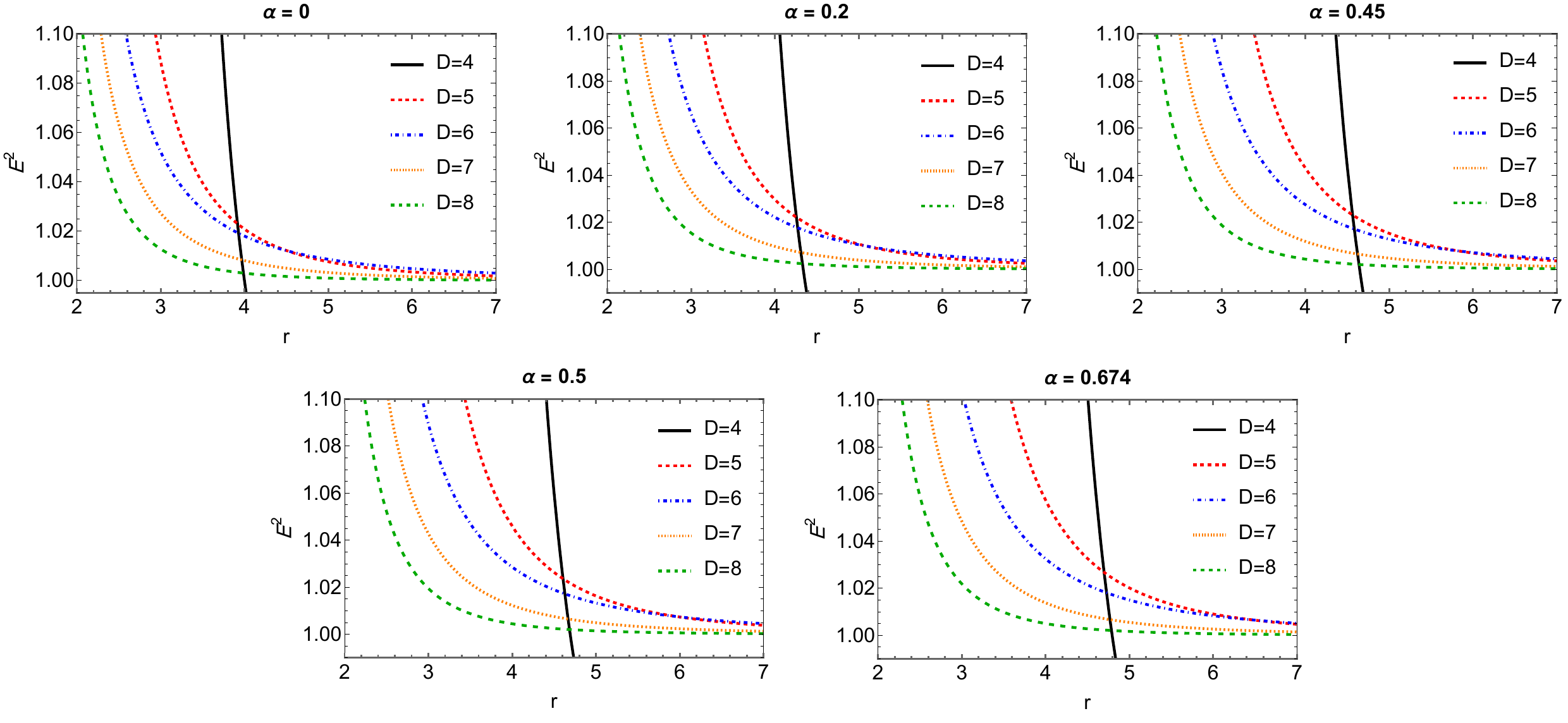}
\label{fig:E-alpha}}
\caption{\small{\emph{Plots of $E^2$ versus $r$ for different $\alpha$ values and spacetime dimensions. (a) illustrates $E^2$ for different values of $\alpha$ with fixed spacetime dimensions, and (b) shows the variation of $E^2$ for different spacetime dimensions with fixed $\alpha$.}}}
\label{fig:E}
\end{figure}

Figure~\ref{fig:E} illustrates the behavior of $E^{2}$ as a function of the radial coordinate $r$ for various values of the STVG parameter $\alpha$ and different spacetime dimensions. The case $\alpha=0$ corresponds to the Schwarzschild spacetime. For a fixed dimension $D$ [see Fig.~\ref{figE-D}], $E^{2}$ increases monotonically with larger values of $\alpha$. In contrast, the influence of $\alpha$ on $E^{2}$ becomes less pronounced in higher dimensions. On the other hand, for a given value of $\alpha$ [see Fig.~\ref{fig:E-alpha}], $E^{2}$ decreases as the number of spacetime dimensions increases, with the rate of decrease being more significant for larger values of $\alpha$.  

\begin{figure}[htb]
\centering
\subfloat[\small{\emph{}}]{\includegraphics[width=1\textwidth]{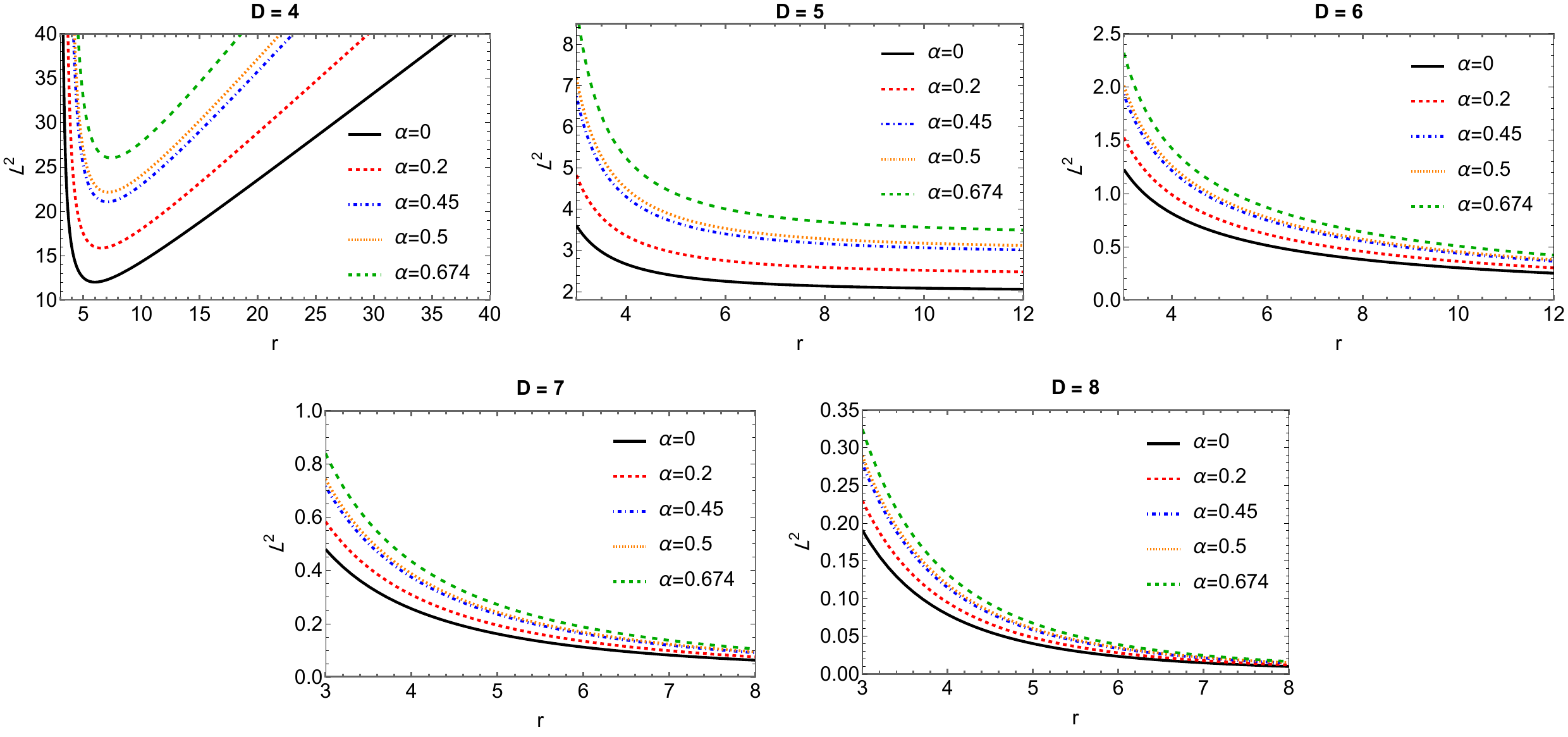}
\label{fig:L-D}}

\subfloat[\small{\emph{}}]{\includegraphics[width=1\textwidth]{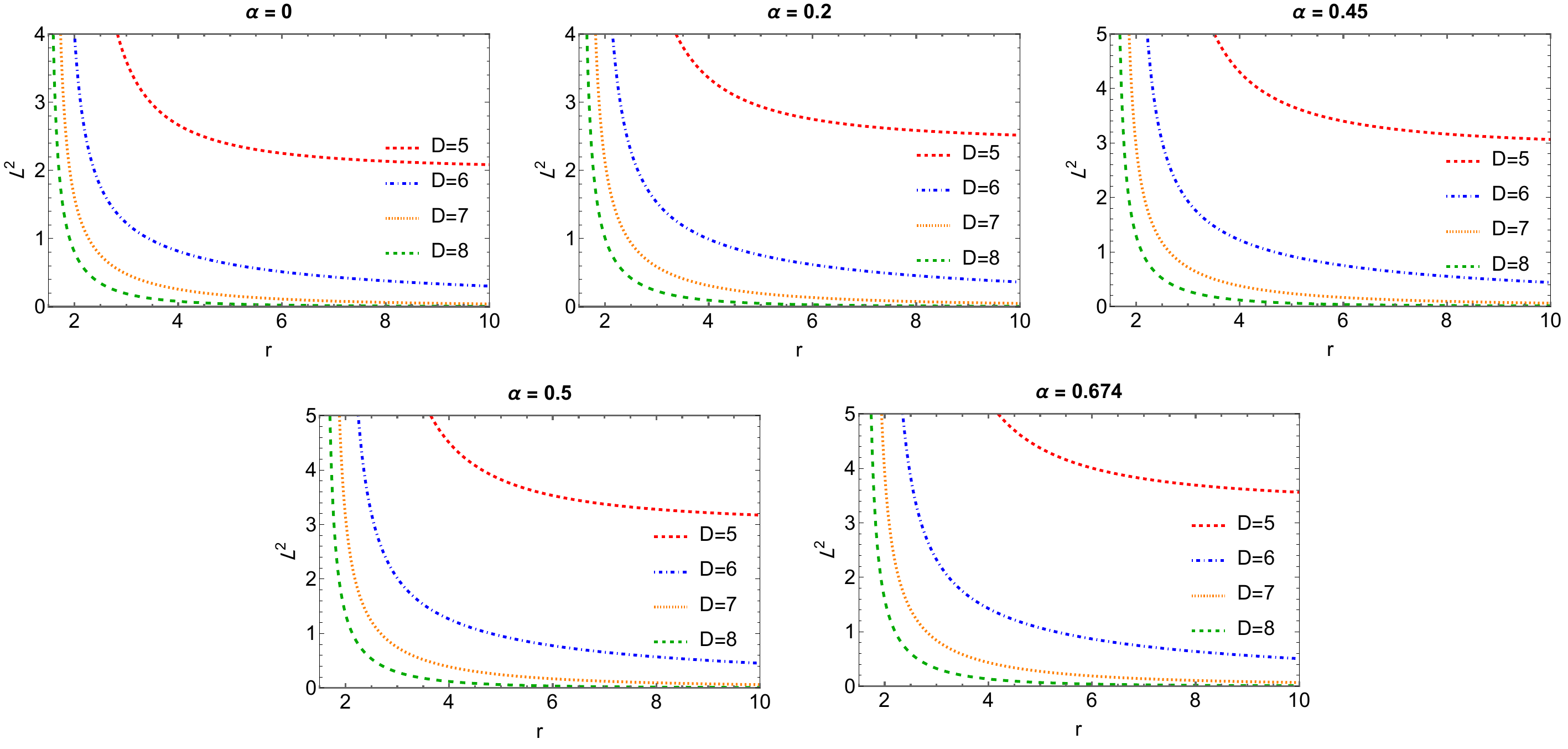}
\label{fig:L-alpha}}
\caption{\small{\emph{Plots of $L^2$ versus $r$ for different $\alpha$ values and spacetime dimensions. (a) illustrates $L^2$ for different values of $\alpha$ with fixed spacetime dimensions, and (b) shows the variation of $L^2$ for different spacetime dimensions with fixed $\alpha$.}}}
\label{fig:L}
\end{figure}

Fig. \ref{fig:L}  shows $L^{2}$ in terms of $r$ for different values of $\alpha$ and $D$. Similar to the behavior of $E^{2}$, in fixed spacetime dimensions (\ref{fig:L-D}), $L^{2}$ increases with increasing $\alpha$. With increasing the spacetime dimensions from $D=4$ to $D=5$, the behavior of  $L^2$ undergoes a fundamental change. Since the gravitational potential changes as $V_{G}\sim1/r^{D-3}$, the gravitational force decreases more rapidly with radius in higher dimensions, which means a particle needs less angular momentum to maintain equilibrium in circular orbits. Therefore, the plots (\ref{fig:L-alpha}) show a significant decrease in $L^2$ in higher dimensions.

\begin{figure}[htb]
\centering
\subfloat[\small{\emph{}}]{\includegraphics[width=1\textwidth]{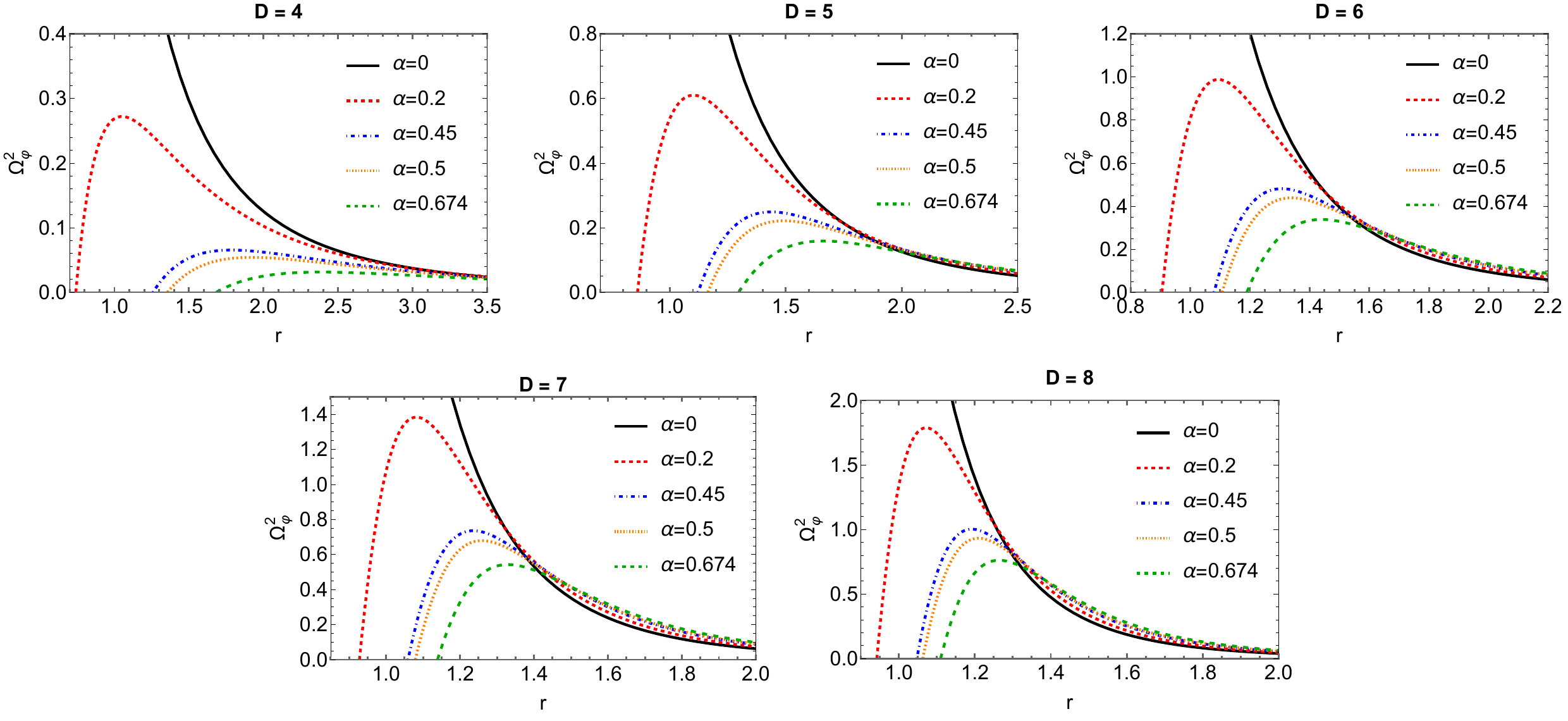}
\label{fig:Omega-D}}

\subfloat[\small{\emph{}}]{\includegraphics[width=1\textwidth]{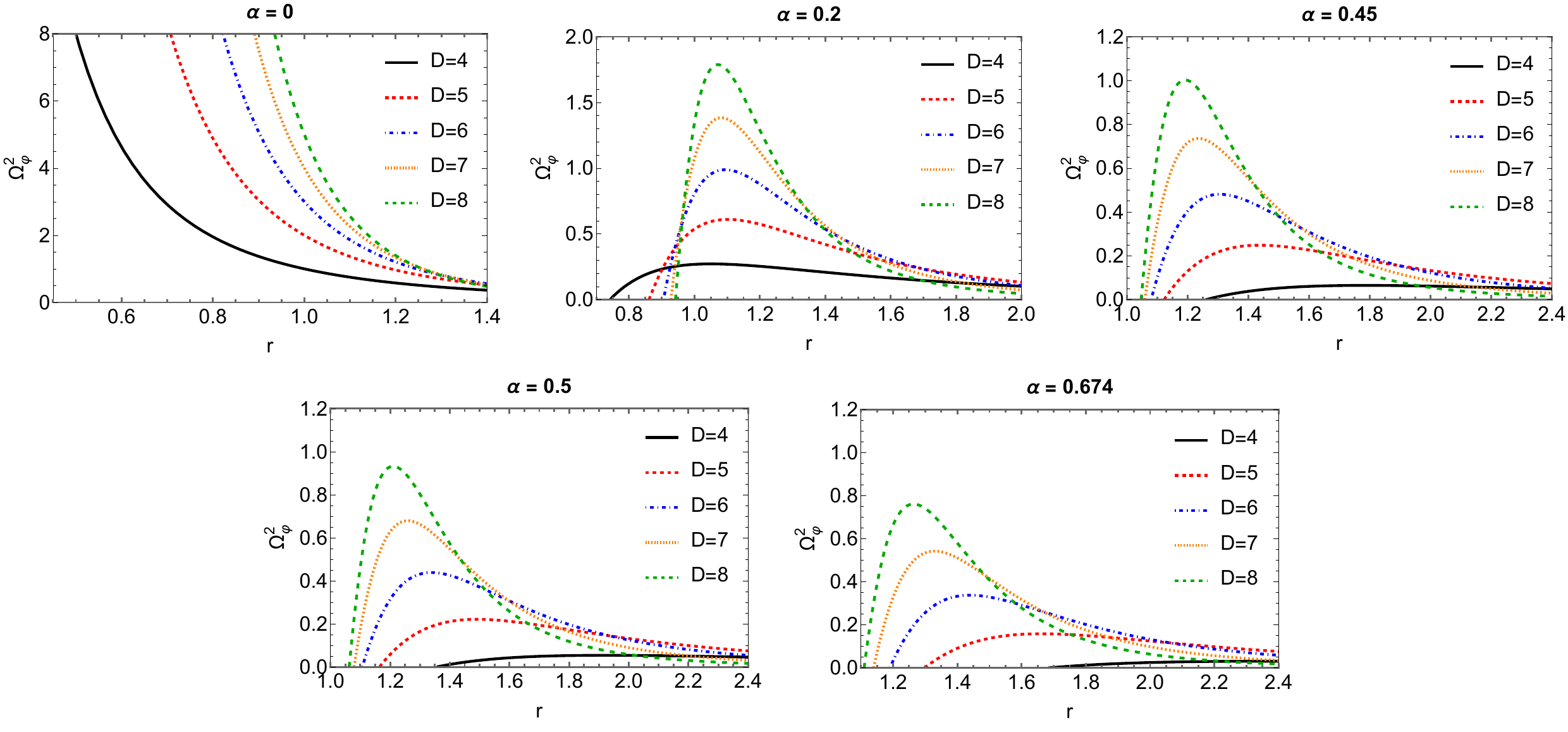}
\label{fig:Omega-alpha}}
\caption{\small{\emph{Plots of ${{\Omega_\varphi}}^2$ versus $r$ for different $\alpha$ values and spacetime dimensions. (a) illustrates ${\Omega_{\varphi}}^2$ for different values of $\alpha$ with fixed spacetime dimensions, and (b) shows the variation of ${\Omega_{\varphi}}^2$ for different spacetime dimensions with fixed $\alpha$.}}}
\label{fig:Omega}
\end{figure}

Figure~\ref{fig:Omega}, plotted for $\Omega_{\varphi}^{2}$, shows that at a given $D$ (\ref{fig:Omega-D}), increasing the value of $\alpha$ up to a certain radius causes the value of $\Omega_{\varphi}^{2}$ to decrease, but after that, $\Omega_{\varphi}^{2}$ increases. As the radius decreases, the angular velocity of the particle relative to the distant observer increases. However, as the particle approaches the event horizon, repulsion force resulting from parameter $\alpha$, causing the angular velocity relative to the external observer to tend towards zero. The values of angular velocity at a fixed $\alpha$ (\ref{fig:Omega-alpha}) initially increases and then decreases with increasing the spacetime dimensions. The rate of these changes decreases with increasing $\alpha$. For $\alpha=0$ case, due to the singularity at the origin, the angular velocity diverges as $r\to0$. At large radial distances from the gravitational source,  $E^{2}$, $L^{2}$, and  $\Omega_{\varphi}^{2}$ tend towards a constant value. We set $M=1$ for plotting the diagrams, so that all quantities are expressed in units of mass.

As we mentioned, the extrema of the effective potential give the radii of circular orbits. The second derivative of the effective potential determines the stability of an orbit. For a stable circular orbit, the effective potential has a minimum, which means the second derivative must be positive, while a negative second derivative indicates a maximum value for the effective potential which yields an unstable circular orbit. Therefore, the existence of the innermost stable circular orbit (ISCO) requires the following conditions
\begin{equation}\label{iscocons}
\frac{dV_{eff}}{dr}=0\,,\qquad \frac{d^{2}V_{eff}}{dr^{2}}=0\,,
\end{equation}
to be satisfied. $r_{_{ISCO}}$ acts as a boundary between stable and unstable circular orbits. Since most of the radiation emitted by the accretion disk comes from its inner regions, the position of $r_{_{ISCO}}$ is one of the most significant factors in determining the total radiative output. $r_{_{ISCO}}$ determines which regions from the accretion disk contribute to the radiation observed at infinity and how the radiation is deflected under the influence of the gravitational field. As a result, $r_{_{ISCO}}$ directly affects the structure and luminosity of the accretion disk and both the size and shape of the black hole's shadow.

Since the equations related to $r_{_{ISCO}}$ become very complicated due to the complex shape of the metric function, only the results obtained from the numerical solutions of the equations using Wolfram Mathematica (v14.2) are given in this section. Table \ref{ISCO} shows $r_{_{ISCO}}$ for a test particle in the spacetime of a higher-dimensional MOG dark compact object. The corresponding results are obtained for two different ranges of $\alpha$. Since the gravitational force decreases rapidly with increasing spacetime dimensions, it is insufficient to balance the centrifugal force even near the central object. An increase in the values of $\alpha$ enhances the gravitational field. Therefore, for spacetime dimensions higher than $D = 4$, we use $\alpha\ge0.674$  to obtain $r_{ISCO}$. As summarized in Table~\ref{ISCO}, the radius of the innermost stable circular orbit 
$r_{_{\text{ISCO}}}$ exhibits two distinct trends: it decreases with increasing spacetime 
dimensionality $D$, and it increases with the MOG parameter $\alpha$. The reduction with $D$ 
arises because the higher-dimensional MOG dark compct object potential falls off more 
steeply, thereby diminishing the radial region where centrifugal 
repulsion can balance the stronger gravitational attraction, which ultimately leads to the 
reduction of stable circular orbits. In contrast, larger 
values of $\alpha$ effectively enhance the gravitational charge term in the metric, which 
shifts the photon sphere and ISCO outward, resulting in a larger stable orbital radius 
compared with the general relativistic case in the same dimension.

\begin{table}[ht]
  \centering
  \caption{\label{Table}\small{\emph{The numerical values of $r_{_{ISCO}}$ for a test particle moving in the regular static spherically symmetric higher-dimensional MOG dark compact object spacetime for various values of $\alpha$ and spacetime dimensions. (a) and (b) show the values of $r_{_{ISCO}}$ for $D=4$ and $D=5,6,7,8$ respectively.}}}
  {\renewcommand{\arraystretch}{1.3}
  \subfloat[\label{tab:a}]{\setlength{\tabcolsep}{8pt}\begin{tabular}{|c||c|c|c|c|c|}\hline
  $\alpha$ & $0$ & $0.2$ & $0.45$ & $0.5$ & $0.674$ \\\hline\hline
  $r_{_{ISCO}} (D = 4)$ & $ 6 $ & $ 6.534 $ & $ 7.06145 $ & $ 7.14706 $ & $ 7.38602 $ \\\hline
  \end{tabular}}
  \\
  \subfloat[\label{tab:b}]{\setlength{\tabcolsep}{8pt}\begin{tabular}{|c||c|c|c|c|}\hline
  $\alpha$ & $r_{_{ISCO}} (D = 5)$ & $r_{_{ISCO}} (D = 6)$ & $r_{_{ISCO}} (D = 7)$ & $r_{_{ISCO}} (D = 8)$ \\\hline\hline
  $0.674$ & $1.28557$ & $1.18228$ & $1.1338$ & $1.10567$ \\\hline
  $1$ & $1.34844$ & $1.21595$ & $1.15619$ & $1.12227$ \\\hline
  $1.5$ & $1.53973$ & $1.32509$ & $1.23191$ & $1.18007$ \\\hline
  $2$ & $1.71619$ & $1.42269$ & $1.29866$ & $1.23061$ \\\hline
  $2.45$ & $1.86192$ & $1.50106$ & $1.35155$ & $1.27035$ \\\hline
  \end{tabular}}}
  \label{ISCO}
\end{table}

\subsection{Radiant energy flux}\label{energy flux}

Infalling matter accreting onto a dark compact object releases gravitational energy in the form of electromagnetic radiation. This radiation originates from the accretion disk, and its distribution is described by the radiant energy flux. The flux profile depends on the motion of particles in the accretion disk, characterized by their specific energy, specific angular momentum, and angular velocity, which are directly influenced by the spacetime geometry. The expression for the radiant energy flux given in \cite{Kato:20008,Page1974} takes the following form for our $D$-dimensional case
\begin{equation}\label{radflu1}
\mathcal{F}(r)=-\frac{\dot{M}}{\omega_{_{D-2}}}\frac{\Omega'_{\varphi}}{\sqrt{-g}\left(E-L\Omega_{\varphi}\right)^{2}}\int_{r_{_{ISCO}}}^{r}\left(E-L\Omega_{\varphi}\right)L'dr\,.
\end{equation}

\begin{figure}[htb]
\centering
\subfloat[\label{Fa} $D=4$]{\includegraphics[width=0.475\textwidth]{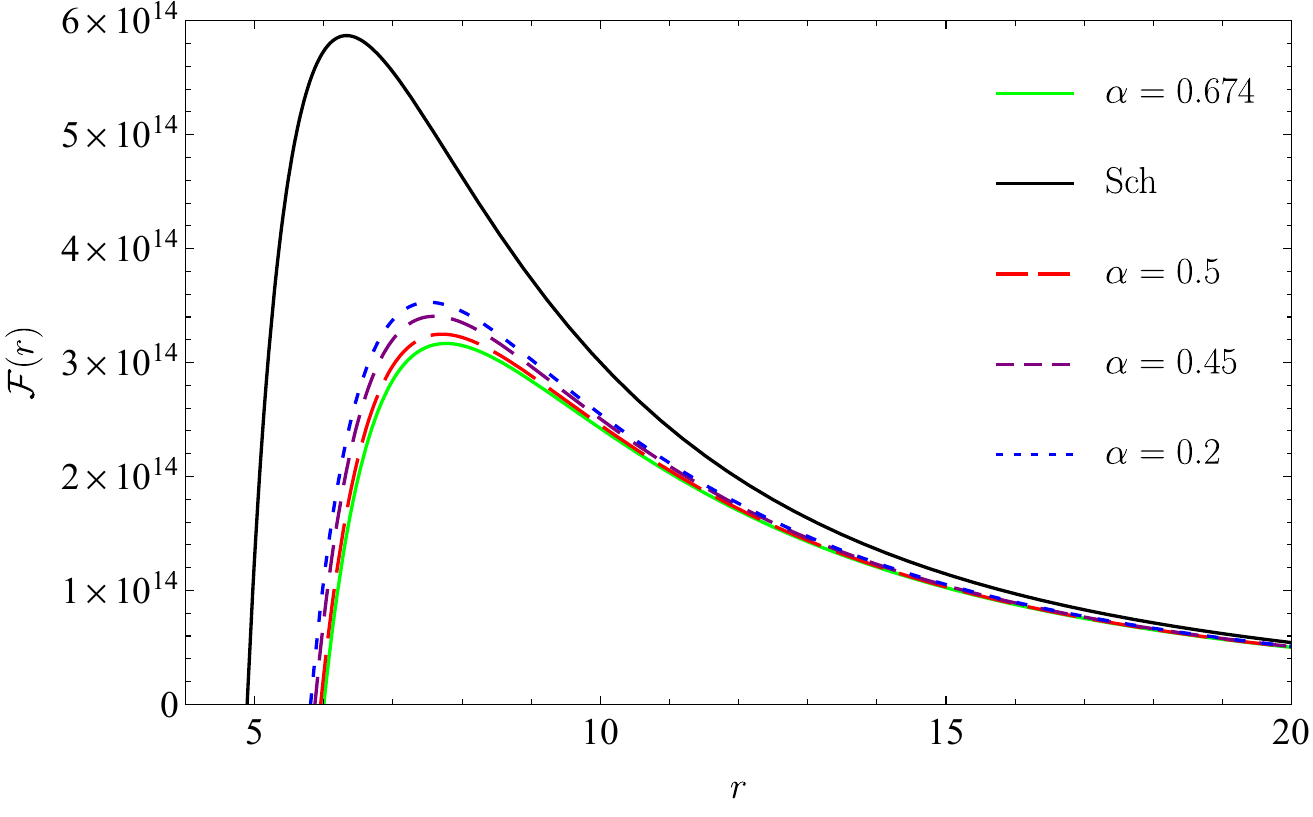}}
\,\,\,
\subfloat[\label{Fb} $D=5$]{\includegraphics[width=0.475\textwidth]{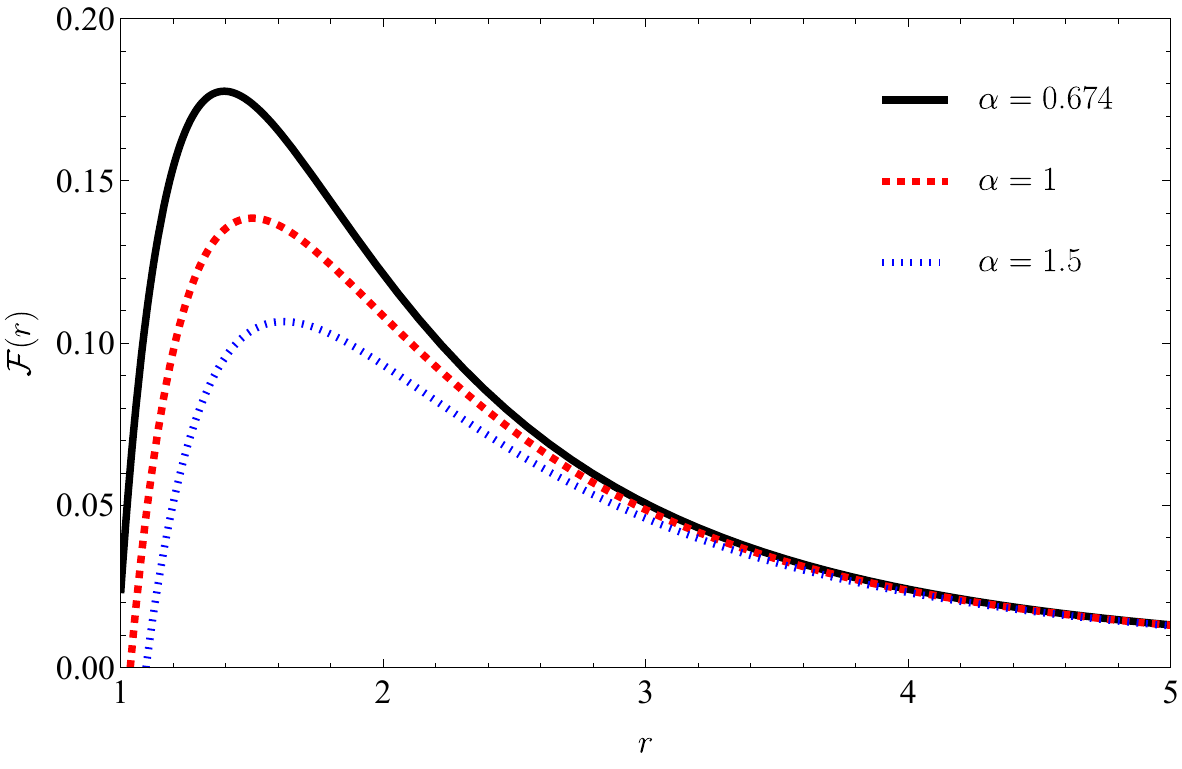}}
\caption{\label{F}\small{\emph{The behavior of $\mathcal{F}(r)$ versus $r$ for different values of $\alpha$ at $D=4$ and $D=5$. For the case $D=5$, the values of $\mathcal{F}(r)$ are multiplied by ${10}^{16}$.}}}
\end{figure}

Figure~\ref{F} illustrates the radial dependence of the energy flux $\mathcal{F}(r)$ emitted 
by the accretion disk of the higher dimensional MOG dark compact object. For $r > r_{_{\text{ISCO}}}$, the flux 
initially rises, attaining a maximum near the inner disk region, and subsequently decreases 
monotonically toward zero at large radii, where the gravitational binding energy available 
for radiation becomes negligible. The peak value of $\mathcal{F}(r)$ decreases 
as $\alpha$ increases. In contrast, increasing the number of spacetime dimensions $D$ steepens 
the gravitational potential, which enhances the binding energy released during accreting
and thus leads to a higher peak in the flux profile. 

For clarity, Figure~\ref{Fa} displays the flux profile in the four-dimensional case ($D=4$), 
which reproduces the general relativistic behavior modified by $\alpha$. Figure~\ref{Fb}, on 
the other hand, corresponds to the five-dimensional case ($D=5$), where the peak flux is 
shifted upward relative to $D=4$, reflecting the stronger gravitational field in higher 
dimensions. A systematic comparison between these figures highlights the competing effects of 
$\alpha$ and $D$: while $\alpha$ suppresses the energy flux of accretion disk for higher dimensional MOG dark compact object, 
higher $D$ enhances the energy flux due to the more rapid falloff of the ISCO radius.

\begin{figure}[htb]
\centering
\subfloat[\label{Ta} $D=4$]{\includegraphics[width=0.475\textwidth]{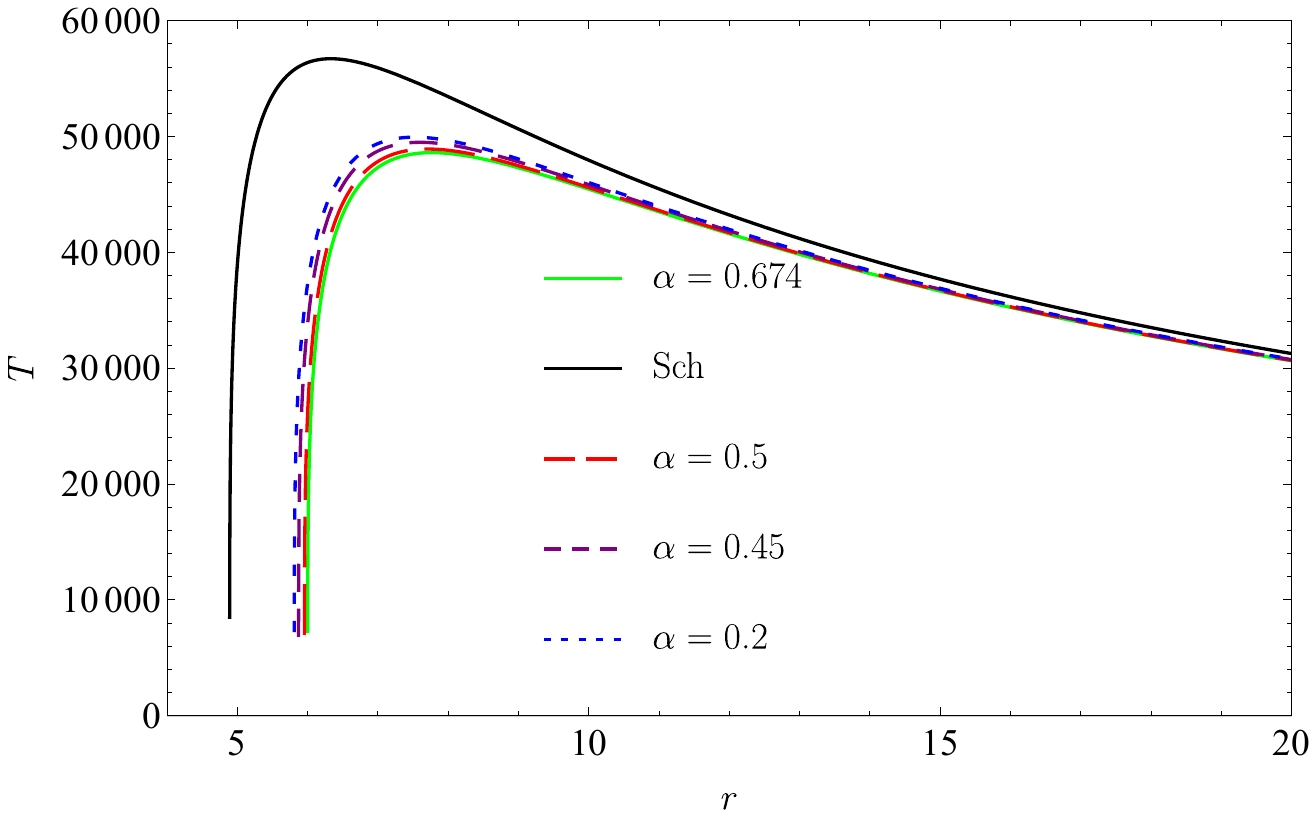}}
\,\,\,
\subfloat[\label{Tb1} $D=5$]{\includegraphics[width=0.475\textwidth]{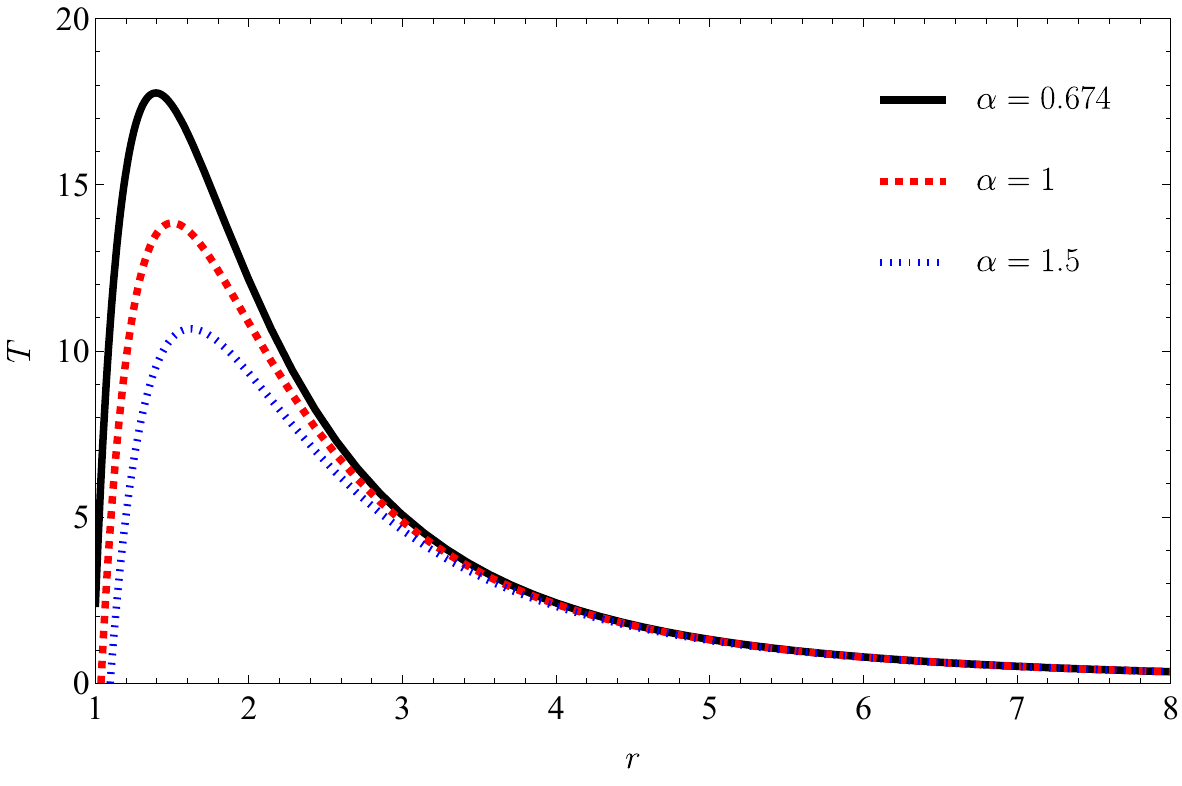}}
\caption{\label{T}\small{\emph{The behavior of $T_{eff}$ versus $r$ for different values of $\alpha$ at $D=4$ and $D=5$. For the case $D=5$, the values of $T$ are multiplied by ${10}^{5}$.}}}
\end{figure}

Assuming thermodynamical equilibrium in the accretion disk, the radiation emitted from its surface can be approximated as black body radiation \cite{Perez:2017spz,Kato:20008,Page1974}. In this case, the effective temperature of the disk can be obtained using the Stefan-Boltzmann law $\mathcal{F}(r)=\sigma_{_{SB}}T_{eff}^{4}$, in which $\sigma_{_{SB}}$ is Stefan-Boltzman constant. Figure \ref{T} illustrates effective temperature $T_{eff}$ versus $r$ for $D=4$ (Fig. \ref{Ta}) and $D=5$ (Fig. \ref{Tb1}). As can be seen, the temperature of the accretion disk of higher dimensional MOG dark compact object behaves similarly to the energy flux. Overall, the effective temperature profile of the higher dimensional of MOG dark compact objcet reflects the competition between $\alpha$ and $D$: larger 
$\alpha$ leads to a cooler disk, while higher $D$ yields a hotter disk with a more pronounced 
inner--disk emission.

\begin{figure}[htb]
\centering
\subfloat[\label{La} $D=4$]{\includegraphics[width=0.475\textwidth]{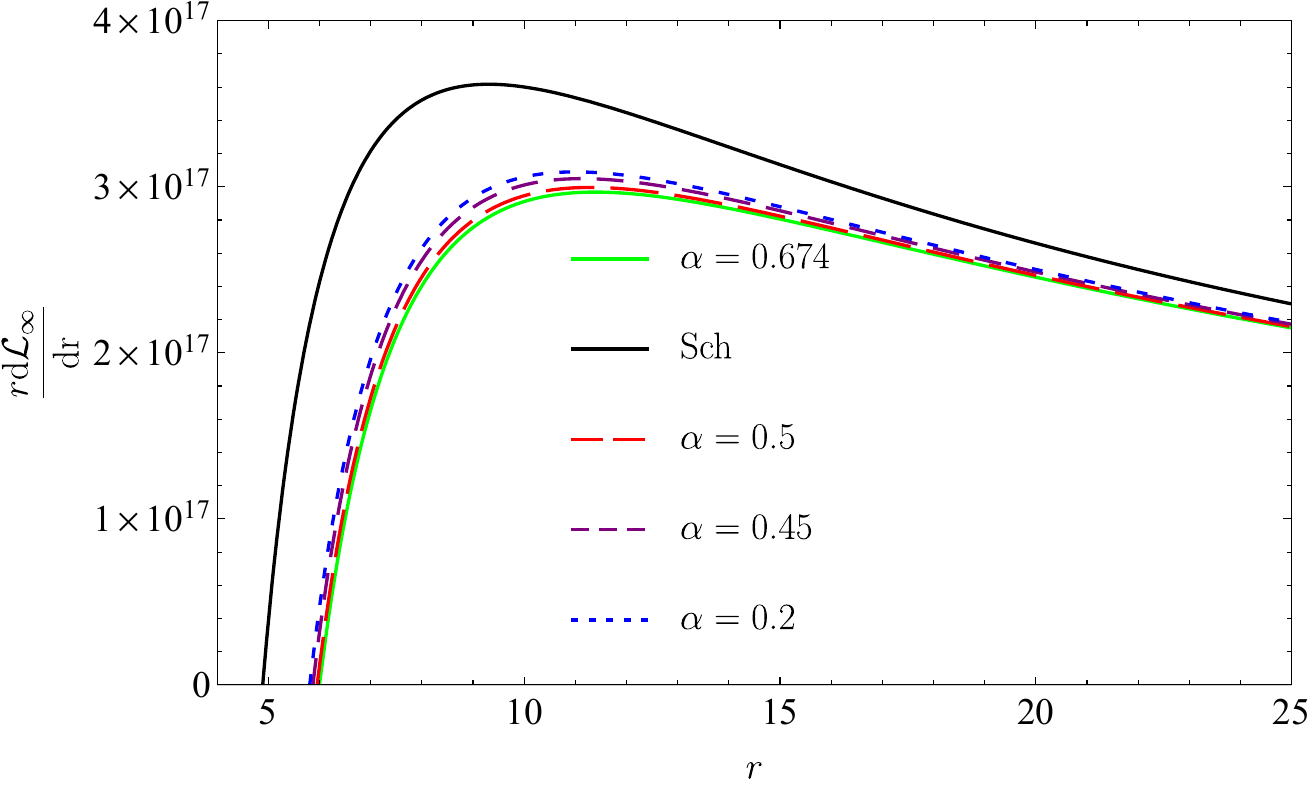}}
\,\,\,
\subfloat[\label{Lb} $D=5$]{\includegraphics[width=0.475\textwidth]{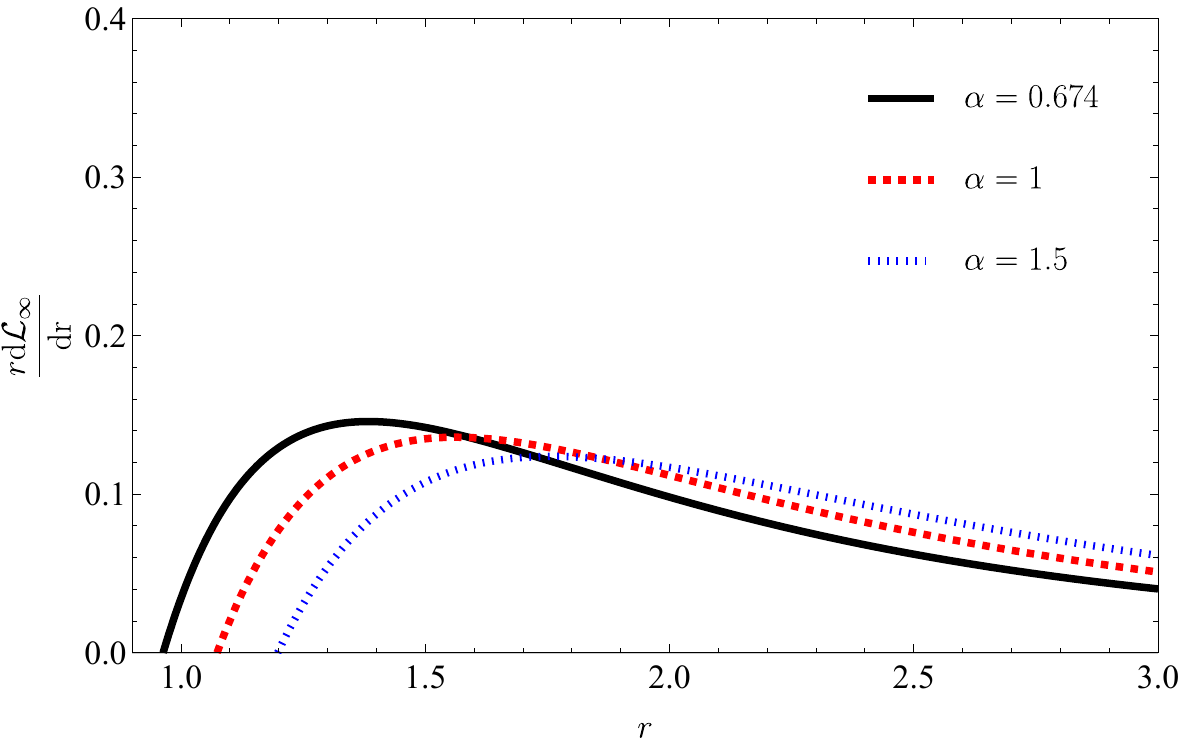}}
\caption{\label{L}\small{\emph{The behavior of $\frac{rd\mathcal{L}_\infty}{dr}$ versus $r$ for different values of $\alpha$ at $D=4$ and $D=5$. For the case $D=5$, the values of $\frac{rd\mathcal{L}_\infty}{dr}$ are multiplied by ${10}^{19}$.}}}
\end{figure}

Another key observable for distant detection is the differential luminosity measured by an 
observer at infinity, which is defined as \cite{novikov1973astrophysics,shakura1973black} 
\begin{equation}\label{Eq:luminosity}
\frac{d\mathcal{L}_{\infty}}{d\ln r} = 4\pi r \sqrt{-g}\,E\,\mathcal{F}(r)\, .
\end{equation}
We compute the differential luminosity of the accretion disk of higher dimensional MOG dark compact object and display its radial dependence 
in Fig.~\ref{L}. The overall behavior resembles that of the flux profile: the luminosity 
rises sharply outside the ISCO, attains a peak in the inner disk region, and gradually declines 
at larger radii, where the available binding energy diminishes. The MOG parameter $\alpha$ 
plays a central role in shaping this profile. As $\alpha$ increases, the reduced specific 
energy release between adjacent circular orbits suppresses the flux and, consequently, lowers 
the differential luminosity. Conversely, higher spacetime dimensions $D$ steepen the 
gravitational potential and enhance the efficiency of energy extraction, which results in a 
brighter disk with a more pronounced luminosity peak near the MOG dark compact object. This demonstrates 
that while $\alpha$ acts to reduce the radiative output of the disk, increasing $D$ tends to 
intensify it, with the strongest impact localized in the inner regions close to the compact 
object.

\subsection{Confrontation with observation}

Following Refs.~\cite{akiyama2022first,doeleman2008event,raymond2024first}, we now turn to 
the question of whether the modifications introduced by the MOG parameter $\alpha$ and the 
spacetime dimensionality $D$ can lead to signatures accessible to current observations. In 
particular, we quantify how the presence of $\alpha$ and higher dimensions alters the corresponding effective temperature profile. These theoretical deviations 
are then evaluated against the observational capabilities of present very long baseline 
interferometry (VLBI) facilities, with emphasis on the Event Horizon Telescope (EHT). To this 
end, it is convenient to introduce the relative variations in the temperature, which 
serve as direct measures of the observational impact of MOG dark compact objects.

\begin{equation}\label{delta}
\delta T_{\mathrm{eff}}:=\frac{T_{\mathrm{eff}}^{(\mathrm{STVG}_{D})}-T_{\mathrm{eff}}^{\mathrm{(Sch)}}}{T_{\mathrm{eff}}^{\mathrm{(Sch)}}} ,
\end{equation}
in which ($\mathrm{STVG}_{D}$) and (Sch) refer to higher-dimensional STVG theory case and Schwarzschild case respectively. EHT does not directly measure the effective temperature of the accretion disk. Rather, millimeter-wavelength VLBI observations, such as those conducted by the EHT, are capable of 
reconstructing the brightness distribution and overall image morphology of compact objects. 
The data are characterized by an angular resolution on the order of $20$--$25\,\mu$as at an 
observing frequency of $230\,\mathrm{GHz}$ ($\lambda = 1.3\,\mathrm{mm}$), with a typical 
uncertainty in the measured brightness temperature of about $10$--$20\%$ under optimal 
conditions. The relationship between the luminosity $I(\nu)$ and the corresponding brightness temperature $T_{b}$, which EHT reconstruct from data,  is given by
\begin{equation}\label{Tb}
T_b(\nu)=\frac{c^2}{2k_{\mathrm{B}}\nu^2}\,I_{\nu}(\nu)
\end{equation}
in which $k_{\mathrm{B}}$ is Boltzmann constant.

In general, the brightness temperature is not equal to the effective temperature of the surface. However, if the accretion disk radiates as a blackbody (in the Novikov-Thorne model), in the long-wavelength regime corresponding to the Rayleigh-Jeans approximation relevant for 
the EHT observing frequency, 
the brightness temperature obtained from the Eq. \eqref{Tb} can be used to approximate the effective temperature of the disk.
The results obtained for $\delta T_{\mathrm{eff}}$ for different values of the parameter $\alpha$ around $r_{_{ISCO}}$ at $D=4$ and $D=5$ are shown in Table \ref{FTL}.

\begin{table}[h!]
  \centering
  \caption{\label{Table}\small{\emph{Values of the relative shifts in effective temperature around $r_{_{ISCO}}$ for the different values of $\alpha$ at $D=4$ and $D=5$.}}}
  \renewcommand{\arraystretch}{1.3} % فاصله بین سطرها
  {\setlength{\tabcolsep}{8pt}\begin{tabular}{|c||c|c|}\hline
  \toprule
  $\alpha$ & $\delta T_{\text{eff}} (D=4)$ & $\delta T_{\text{eff}} (D=5)$ \\\hline\hline
  \midrule
  $0.2$ & $18\%$ & $-$ \\\hline
  $0.45$ & $13\%$ & $-$ \\\hline
  $0.5$ & $15\%$ & $-$ \\\hline
  $0.674$ & $14\%$ & $29\%$ \\\hline
  $1$ & $-$ & $23\%$ \\\hline
  $1.5$ & $-$ & $18\%$ \\\hline
  \bottomrule
  \end{tabular}}
  \label{FTL}
\end{table}

According to the EHT report, the brightness calibration uncertainties for Sgr A* are on the order of $10\%$$-$$20\%$ \cite{akiyama2022first}. As we can see from Table \ref{FTL}, we conclude that the predicted MOG and extra-dimensional corrections to the disk temperature may lie at the threshold of detectability with the current observational capabilities of the EHT. For $D=4$ , this holds across all values of $\alpha$, whereas for $D=5$ the effect becomes consistent with EHT data for $\alpha\geq1.5$. These findings highlight the potential significance of future precision measurements of thermal disk spectra or reconstructed brightness distributions.

\section{Accretion onto a higher-dimensional MOG dark compact object}\label{ARMDCO}
In this part, our goal is to derive the fundamental dynamical relations and parameters governing accretion onto the highr dimesnional regular MOG dark compact object, following the approach outlined in Refs.~\cite{Nozari:2020swx,Salahshoor:2018plr}. For this purpose, we focus on a spherically symmetric accretion process restricted to the equatorial plane, $\theta=\pi/2$. Moreover, the accreting matter is considered to be an inflowing perfect fluid onto the higher dimensional regular MOG dark compact object. We aim to investigate the influence of extra dimensions on the four-velocity of the perfect fluid and the corresponding energy density.

The stress-energy tensor for a perfect fluid is given by
\begin{equation}\label{SET}
T^{\mu\nu}=(p+\rho)u^{\mu}u^{\nu}-pg^{\mu\nu}\,,
\end{equation}
where $p$ and $\rho$ are pressure and energy density, respectively. Considering the motion of the particle to be radial and using the normalization condition of the four-velocity $u^{\mu}u_{\mu}=1$, we are able to obtain $u^{t}$ as follows
\begin{equation}\label{ut}
u^{t}=\frac{\sqrt{f(r)+v^{2}}}{f(r)}=\frac{\sqrt{{ 1
- \frac{(D - 2) m r^{2(D - 3)} \omega_{_{D-2}}}{8\pi \left( r^{2(D - 3)} + \frac{(D - 2)^2 m^2 \alpha (1+\alpha) \omega_{_{D-2}}^2}{256 G^2 \pi^2} \right)^{3/2}}
+ \frac{(D - 3)(D - 2) G q^2 r^{2(D - 3)} \omega_{_{D-2}}^2}{32\pi^2 \left( r^{2(D - 3)} + \frac{(D - 2)^2 m^2 \alpha (1+\alpha) \omega_{_{D-2}}^2}{256 G^2 \pi^2} \right)^2}}+v^{2}}}{ 1
- \frac{(D - 2) m r^{2(D - 3)} \omega_{_{D-2}}}{8\pi \left( r^{2(D - 3)} + \frac{(D - 2)^2 m^2 \alpha (1+\alpha) \omega_{_{D-2}}^2}{256 G^2 \pi^2} \right)^{3/2}}
+ \frac{(D - 3)(D - 2) G q^2 r^{2(D - 3)} \omega_{_{D-2}}^2}{32\pi^2 \left( r^{2(D - 3)} + \frac{(D - 2)^2 m^2 \alpha (1+\alpha) \omega_{_{D-2}}^2}{256 G^2 \pi^2} \right)^2}}\,.
\end{equation}

We consider $v<0$ since the accretion is an inward flow of matter. The conservation of the stress-energy tensor $\nabla_\nu T^{\mu\nu}=0$ , gives the following relation
\begin{equation}\label{C0}
(p+\rho)vr^{D-2}\sqrt{ 1
- \frac{(D - 2) m r^{2(D - 3)} \omega_{_{D-2}}}{8\pi \left( r^{2(D - 3)} + \frac{(D - 2)^2 m^2 \alpha (1+\alpha) \omega_{_{D-2}}^2}{256 G^2 \pi^2} \right)^{3/2}}
+ \frac{(D - 3)(D - 2) G q^2 r^{2(D - 3)} \omega_{_{D-2}}^2}{32\pi^2 \left( r^{2(D - 3)} + \frac{(D - 2)^2 m^2 \alpha (1+\alpha) \omega_{_{D-2}}^2}{256 G^2 \pi^2} \right)^2}+v^{2}} \equiv C_{0}\,,
\end{equation}
in which $C_{0}$ is a constant of integration. As has been assumed in \cite{John:2013bqa}, one can define the baryon number density \(n\), and the baryon number flux $J^{\mu\nu}=n u^{\nu}$. So, we can use the conservation law for $J^{\mu\nu}$ in the following form
\begin{equation}\label{J}
\nabla_\nu J^{\mu\nu}=\nabla_\nu (n u^{\nu})=0\,,
\end{equation}
to obtain
\begin{equation}\label{emf2}
\frac{d}{dr}nvr^{D-2}=0\,,
\end{equation}
where we can rewrite it as
\begin{equation}\label{C1}
\rho vr^{D-2} \equiv C_{1}\,
\end{equation}
in which $C_{1}$ is an integration constant. Dividing Eq. \eqref{C0} by Eq. \eqref{C1} yields
\begin{equation}\label{C2}
(\frac{p+\rho}{\rho})\sqrt{1
- \frac{(D - 2) m r^{2(D - 3)} \omega_{_{D-2}}}{8\pi \left( r^{2(D - 3)} + \frac{(D - 2)^2 m^2 \alpha (1+\alpha) \omega_{_{D-2}}^2}{256 G^2 \pi^2} \right)^{3/2}}
+ \frac{(D - 3)(D - 2) G q^2 r^{2(D - 3)} \omega_{_{D-2}}^2}{32\pi^2 \left( r^{2(D - 3)} + \frac{(D - 2)^2 m^2 \alpha (1+\alpha) \omega_{_{D-2}}^2}{256 G^2 \pi^2} \right)^2}+v^{2}} \equiv C_{2}\,,
\end{equation}
where $C_{2}$ is a constant. In the above equation, the expression under the square root approaches unity as \(r \to \infty\). Therefore, the equation can be rewritten as follows
\begin{equation}\label{vr}
(\frac{p+\rho}{\rho})\sqrt{1
- \frac{(D - 2) m r^{2(D - 3)} \omega_{_{D-2}}}{8\pi \left( r^{2(D - 3)} + \frac{(D - 2)^2 m^2 \alpha (1+\alpha) \omega_{_{D-2}}^2}{256 G^2 \pi^2} \right)^{3/2}}
+ \frac{(D - 3)(D - 2) G q^2 r^{2(D - 3)} \omega_{_{D-2}}^2}{32\pi^2 \left( r^{2(D - 3)} + \frac{(D - 2)^2 m^2 \alpha (1+\alpha) \omega_{_{D-2}}^2}{256 G^2 \pi^2} \right)^2}+v^{2}}=\frac{\rho_\infty + p_\infty}{\rho_\infty}\,.
\end{equation}
Consequently, we have
\begin{equation}
C_{2}=\frac{\rho_\infty + p_\infty}{\rho_\infty}\,.
\end{equation}

The matter is assumed to satisfy a linear equation of state of the form \( p = w \rho \), where $w$ is a constant equation of state parameter. Substituting this equation of state into Eq. \eqref{C2} yields
\begin{align}\label{v}
v = \frac{1}{w+1} \Bigg[& C_2^2 - (w+1)^2 \Bigg( 1 - \frac{(D-2) m r^{2(D-3)} \omega_{_{D-2}}}{8 \pi \left( r^{2(D-3)} + \frac{(D-2)^2 m^2 \alpha (1+\alpha) \omega_{_{D-2}}^2}{256 G^2 \pi^2} \right)^{3/2}} \nonumber \\& + \frac{(D-3)(D-2) G q^2 r^{2(D-3)} \omega_{_{D-2}}}{32 \pi^2 \left( r^{2(D-3)} + \frac{(D-2)^2 m^2 \alpha (1+\alpha) \omega_{_{D-2}}^2}{256 G^2 \pi^2} \right)^2}\Bigg)\Bigg]^{1/2}
\end{align}

\begin{figure}[htb]
\centering
\subfloat[\small{\emph{}}]{\includegraphics[width=1\textwidth]{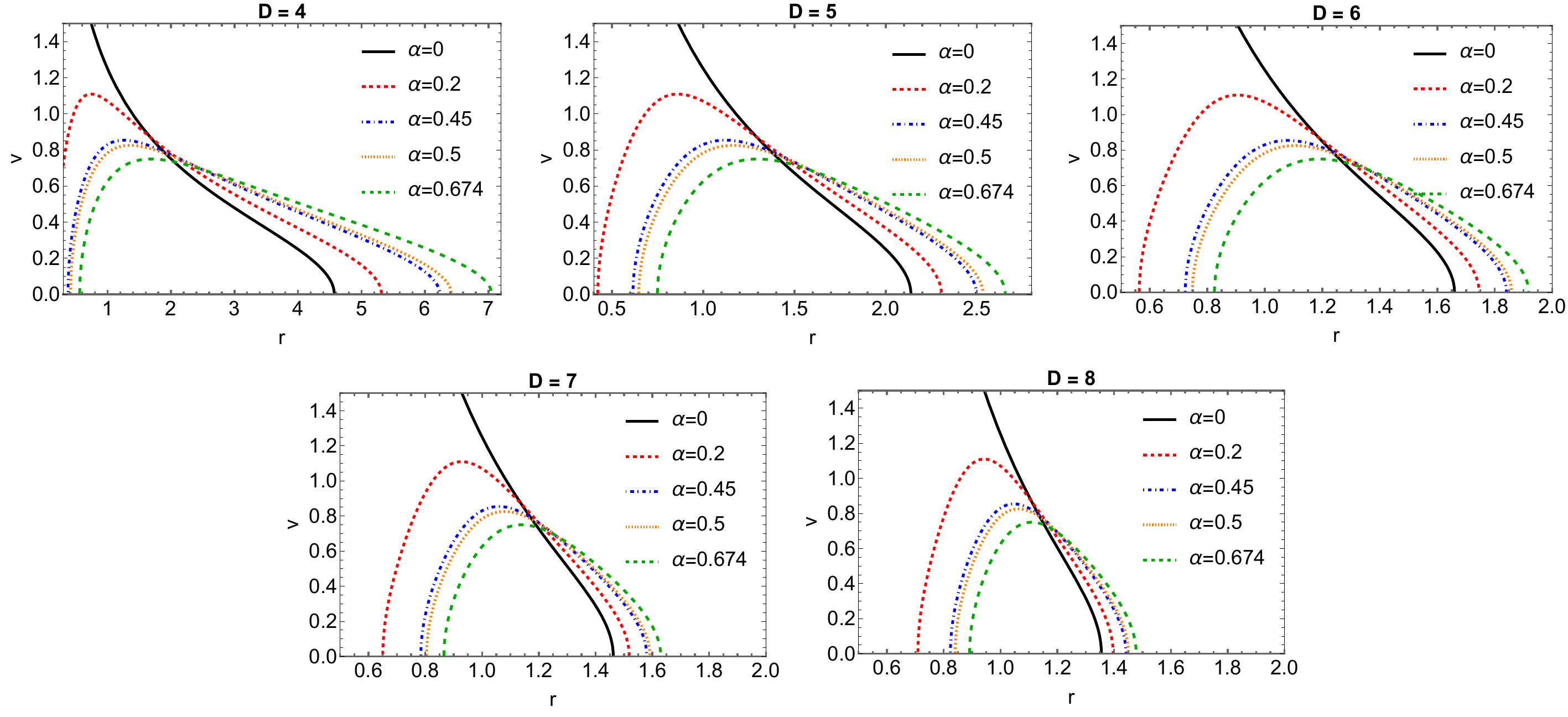}
\label{fig:v-D}}
\hfill
\subfloat[\small{\emph{}}]{\includegraphics[width=1\textwidth]{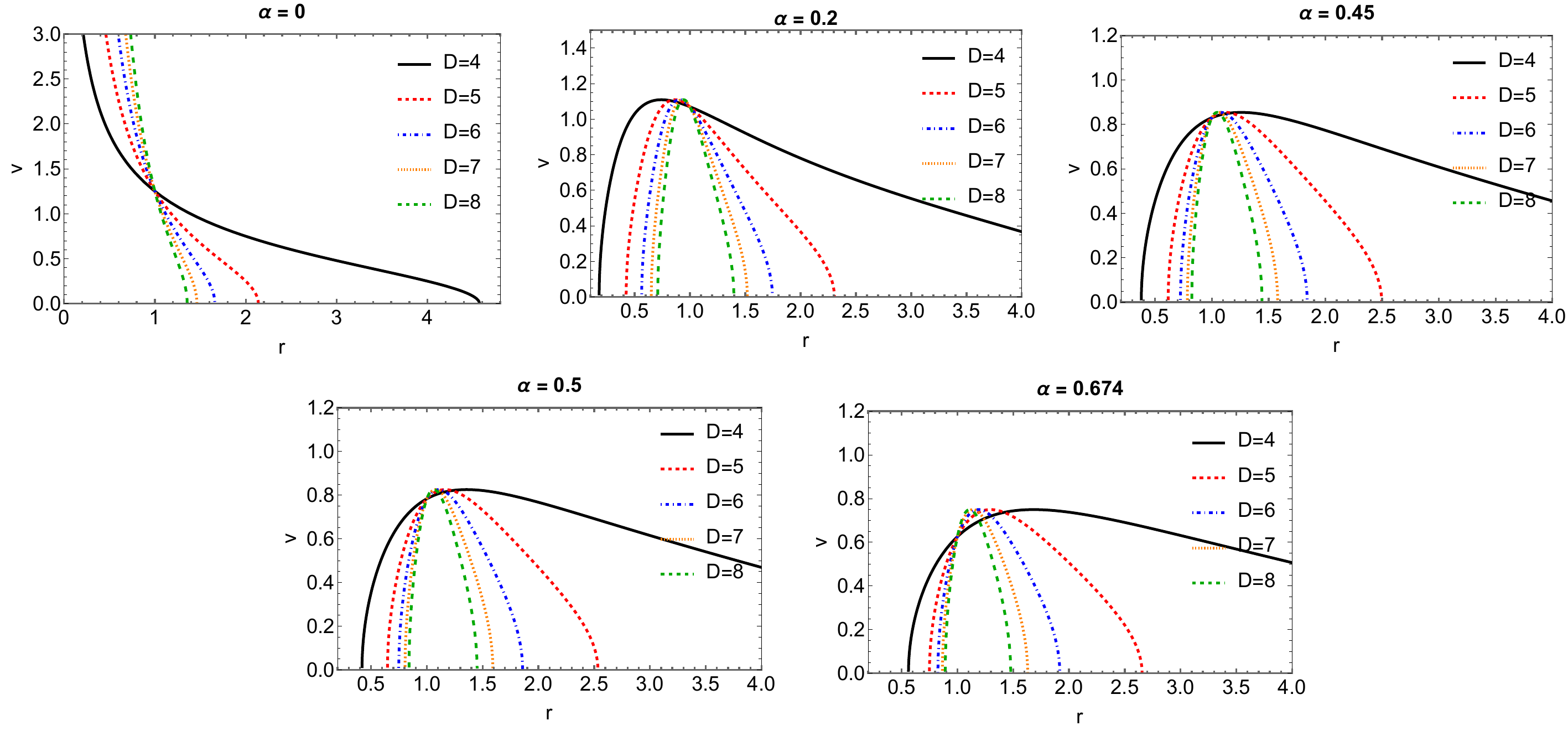}
\label{fig:v-alpha}}
\caption{\small{\emph{The behavior of the radial velocity ($v$) as a function of $r$ is shown for various values of $\alpha$ and the spacetime dimensions. (a) corresponds to varying $\alpha$ with fixed dimension, and (b) corresponds to varying the number of dimensions with fixed $\alpha$.}}}
\label{fig:combined}
\label{fig:v}
\end{figure}

The plots in Figure \ref{fig:v} demonstrate how the radial velocity $v(r)$ of the accreting fluid is shaped by both the MOG parameter $\alpha$ and the number of spacetime dimensions $D$. At large radii, the fluid is initially at rest, but as it approaches the compact object the velocity grows, reaching a peak before gradually decreasing again near the center. This turnover reflects the balance between the inward pull of gravity and the pressure support of the fluid. For the Schwarzschild limit ($\alpha = 0$), the velocity diverges at $r \to 0$ due to the central singularity. By contrast, in the regular MOG case with $\alpha > 0$, the singularity is avoided and the velocity profile remains finite, which is an important physical improvement. Increasing $\alpha$ enhances the effective gravitational attraction, shifting the velocity peak outward and making the flow stronger at intermediate radii. However, raising the dimensionality has the opposite effect: because the gravitational potential scales as $1/r^{D-3}$, gravity decays more rapidly in higher dimensions, which suppresses the radial velocity. This dual influence of $\alpha$ and $D$ highlights how extra dimensions and MOG corrections jointly determine the efficiency of accretion.

From Eqs. \eqref{C1} and \eqref{v} one can derive the following relation for energy density
\begin{equation}\label{rho}
\rho=\frac{C_1}{r^{D-2}}\frac{w+1}{\sqrt{{C_2}^2-\left(w+1\right)^2\left(1-\frac{\left(D-2\right)mr^{2\left(D-3\right)}\omega_{_{D-2}}}{8\pi\left(r^{2\left(D-3\right)}+\frac{\left(D-2\right)^2m^2\alpha\left(1+\alpha\right)\omega_{_{D-2}}^2}{256G^2\pi^2}\right)^{3/2}}+\frac{\left(D-3\right)\left(D-2\right)Gq^2\ r^{2\left(D-3\right)}\omega_{_{D-2}}}{32\pi^2\left(r^{2\left(D-3\right)}+\frac{\left(D-2\right)^2m^2\alpha\left(1+\alpha\right)\omega_{_{D-2}}^2}{256G^2\pi^2}\right)^2}\right)}}\,.
\end{equation}

\begin{figure}[htb]
\centering
\subfloat[\small{\emph{}}]{\includegraphics[width=1\textwidth]{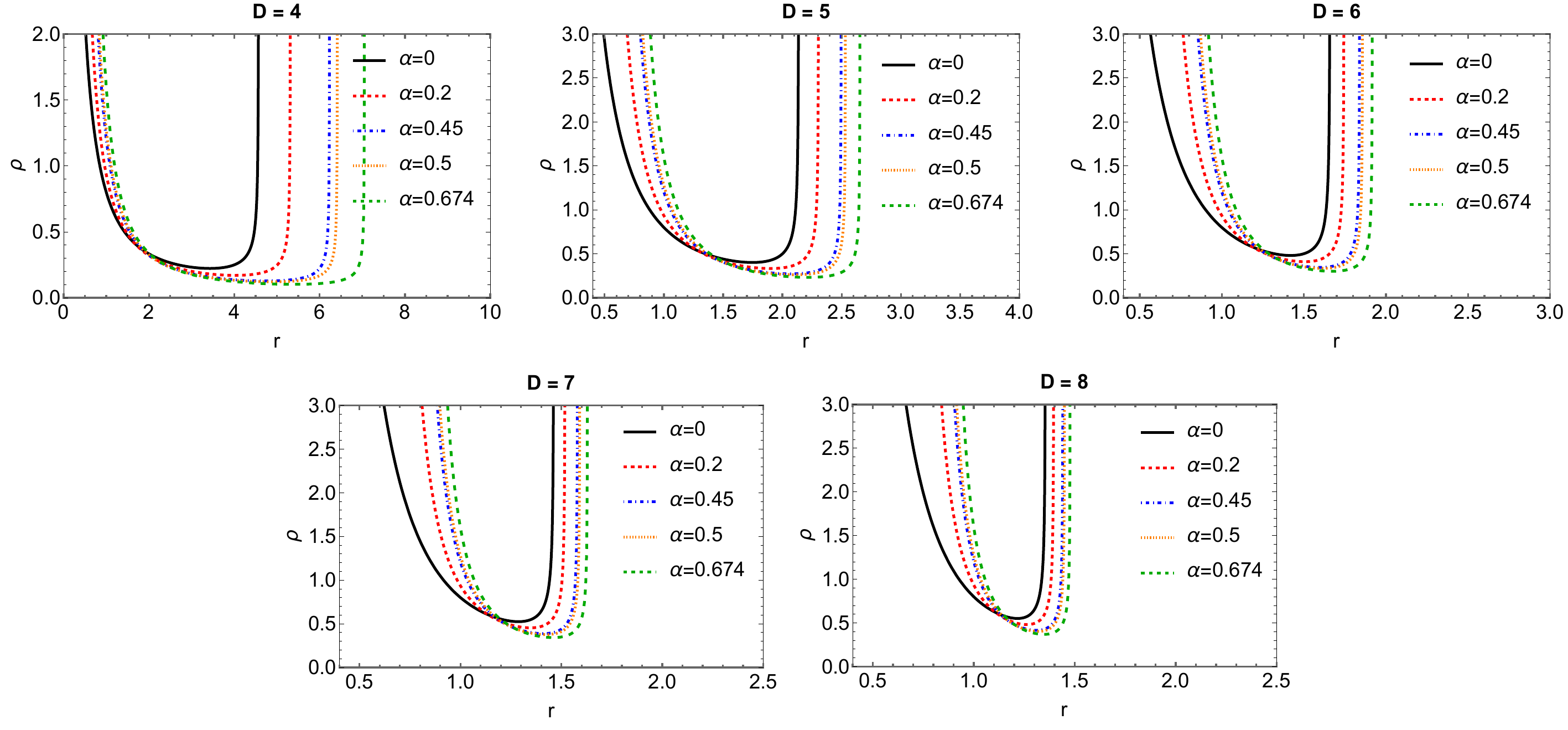}
\label{fig:rho-D}}
\hfill
\subfloat[\small{\emph{}}]{\includegraphics[width=1\textwidth]{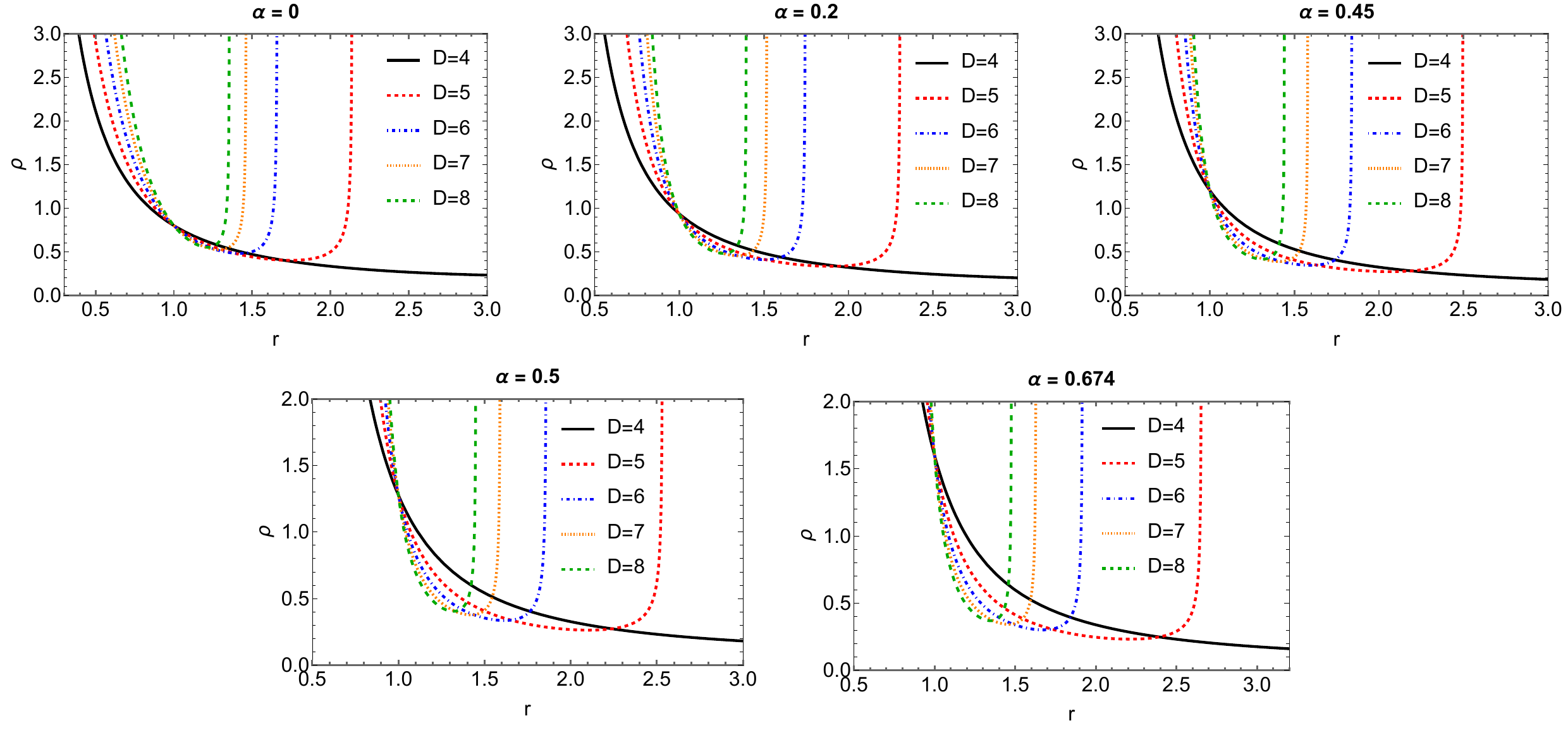}
\label{fig:rho-alpha}}
\caption{\small{\emph{The behavior of energy density ($\rho$) as a function of $r$ is shown for various values of $\alpha$ and spacetime dimensions. (a) corresponds to varying $\alpha$ with fixed dimension, and (b) corresponds to varying the number of dimensions with fixed $\alpha$.}}}
\label{fig:rho}
\end{figure}

The energy density $\rho(r)$ reflects the same competition but from a thermodynamic perspective. As in Figure  \ref{fig:rho}, it diverges near the compact object, reaches a minimum at the location of maximum velocity, and then increases again at large radii as the velocity tends to zero. This non-monotonic behavior is a direct outcome of baryon number conservation and energy flux balance. Larger values of $\alpha$ broaden the density distribution and push the profiles outward, since the enhanced gravitational charge enlarges the effective capture region. On the other hand, increasing the number of dimensions $D$ steepens the falloff of gravity, which accelerates the variation of $\rho(r)$. Compared with the Schwarzschild case, the regular MOG spacetime smooths out the near-center divergence while still allowing strong density gradients close to the object. This suggests that the density profiles of accreting matter could provide indirect observational signatures of both MOG effects and extra dimensions.

\subsection{Mass evolution}

In the process of accretion onto a dark compact object, the mass of the object increases with time. The mass accretion rate is given by the following relation \cite{Michel:1972}
\begin{equation}\label{Mdot}
\dot{M}=-\int T^{r}_{t}dS\,,
\end{equation}
in which $dS=\left(\sqrt{-g}\right)d\theta_{i} d\varphi$ is the surface element of the object and $g=\det(g_{\mu\nu})=-r^{2(D-2)}\prod_{i=1}^{D-3} \sin^2\theta_i\,$ is the determinant of the background metric tensor
associated with the line element \eqref{ds}, which for $\theta_{i}=\frac{\pi}{2}$ takes the form $g=\det(g_{\mu\nu})=-r^{2(D-2)}$. From Eqs. \eqref{SET} and \eqref{ut} the accretion rate $\dot{M}$ can be obtained as
\begin{equation}\label{mdot}
\dot{M}=-\omega_{_{D-2}} r^{D-2}v(p+\rho)\sqrt{1- \frac{(D - 2) m r^{2(D - 3)} \omega_{_{D-2}}}{8\pi \left( r^{2(D - 3)} + \frac{(D - 2)^2 m^2 \alpha (1+\alpha) \omega_{_{D-2}}^2}{256 G^2 \pi^2} \right)^{3/2}}+ \frac{(D - 3)(D - 2) G q^2 r^{2(D - 3)} \omega_{_{D-2}}^2}{32\pi^2 \left( r^{2(D - 3)} + \frac{(D - 2)^2 m^2 \alpha (1+\alpha) \omega_{_{D-2}}^2}{256 G^2 \pi^2} \right)^2}+v^{2}}\,.
\end{equation}

Using Eq. \eqref{C0} one can rewrite Eq. \eqref{mdot} as
\begin{equation}
\dot{M}=-\omega_{_{D-2}} C_{0}\,,
\end{equation}
where $C_{0}$ is a constant defined in Eq. \eqref{C0}.

\subsection{Critical values}

Due to the stronger gravitational field near the central object, the convergence of the flow causes an increase in the density of the fluid. The internal pressure of the fluid resists the increase in density. However, as the velocity of the fluid increases and approaches the sound speed at the critical point, the changes in pressure and density within the fluid become coupled. The critical point is the location where the fluid flow changes from subsonic to supersonic. Identifying critical points is important for studying the dynamic characteristics of the flow and the accretion rate.
The derivatives of Eqs. \eqref{C1} and \eqref{C0} yields
\begin{equation}\label{dC1}
\frac{\rho'}{\rho}+\frac{v'}{v}=-\frac{D-2}{r}\,
\end{equation}
and
\begin{equation}\label{dC0}
\begin{split}
\frac{\rho'}{\rho} \left( \frac{d \ln[p + \rho]}{d \ln[\rho]} - 1 \right) + \frac{v v'}{v^2 + f(r)} + \frac{1}{2}\frac{f'(r)}{v^2 + f(r)} = 0\,,
\end{split}
\end{equation}
respectively. From Eqs. \eqref{dC1} and \eqref{dC0}, one can obtain the following relation
\begin{equation}\label{Ns}
\frac{d\ln[v]}{d\ln[r]}=\frac{\mathcal{N}_{1}}{\mathcal{N}_{2}}\,,
\end{equation}
in which we defined \({\mathcal{N}}_{1}\) and \({\mathcal{N}}_{2}\) as
\begin{equation}\label{N1}
\mathcal{N}_{1}\equiv \frac{r}{2}f^\prime\left(r\right)\left(f\left(r\right)+v^2\right)-\left(D-2\right)K^2\,
\end{equation}
and
\begin{equation}\label{N2}
\mathcal{N}_{2}\equiv K^2-\frac{v^2}{f\left(r\right)+v^2}\,,
\end{equation}
where
\begin{equation}\label{V2}
K^{2}\equiv\frac{d\ln[p+\rho]}{d\ln[\rho]}-1\,.
\end{equation}

The critical point in an accretion disk is a point where the ratio \(\frac{\mathcal{N}_{1}}{\mathcal{N}_{2}}\) approaches the indeterminated form \(\frac{0}{0}\). This condition indicates a change in the behavior of the flow, which corresponds to the transition from subsonic to supersonic. For the equation of motion for the flow to maintain continuity at this point, both the numerator and denominator of the ratio \(\frac{\mathcal{N}_{1}}{\mathcal{N}_{2}}\) must become zero. Therefore, using the condition $\mathcal{N}_{1}=\mathcal{N}_{2}=0$ yields the expression for \(K\) at the critical point in the following form
\begin{equation}\label{Vc2}
K_c^2 = \frac{r f'(r)}{r f'(r) + 2(D - 2) f(r)}\,,
\end{equation}
where the index ($c$) stands for critical values. Since we have $f'(r)>0$ outside the event horizon, the denominator of the expression must be positive. This requirement leads to the following inequality
\begin{equation}\label{racrra}
r f'(r) + 2(D - 2) f(r)>0\,.
\end{equation}
Also, using the condition defining critical point results in
\begin{equation}\label{vc2}
\begin{split}
v_{c}^{2} & =\frac{rf^\prime\left(r\right)}{2\left(D-2\right)}\,.
\end{split}
\end{equation}

\begin{figure}[htb]
\centering
\subfloat[\small{\emph{}}]{\includegraphics[width=0.9\textwidth]{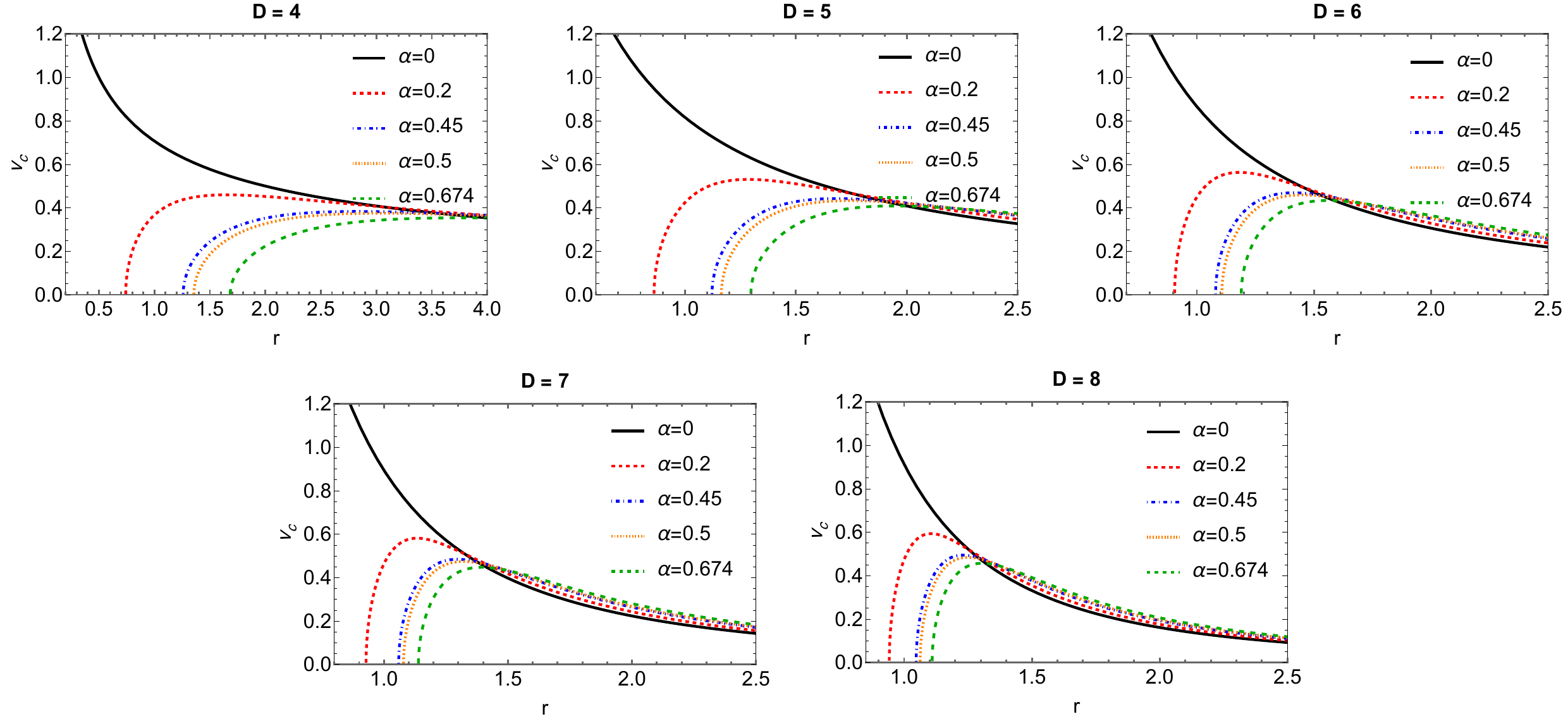}
\label{fig:vc-D}}
\hfill
\subfloat[\small{\emph{}}]{\includegraphics[width=0.9\textwidth]{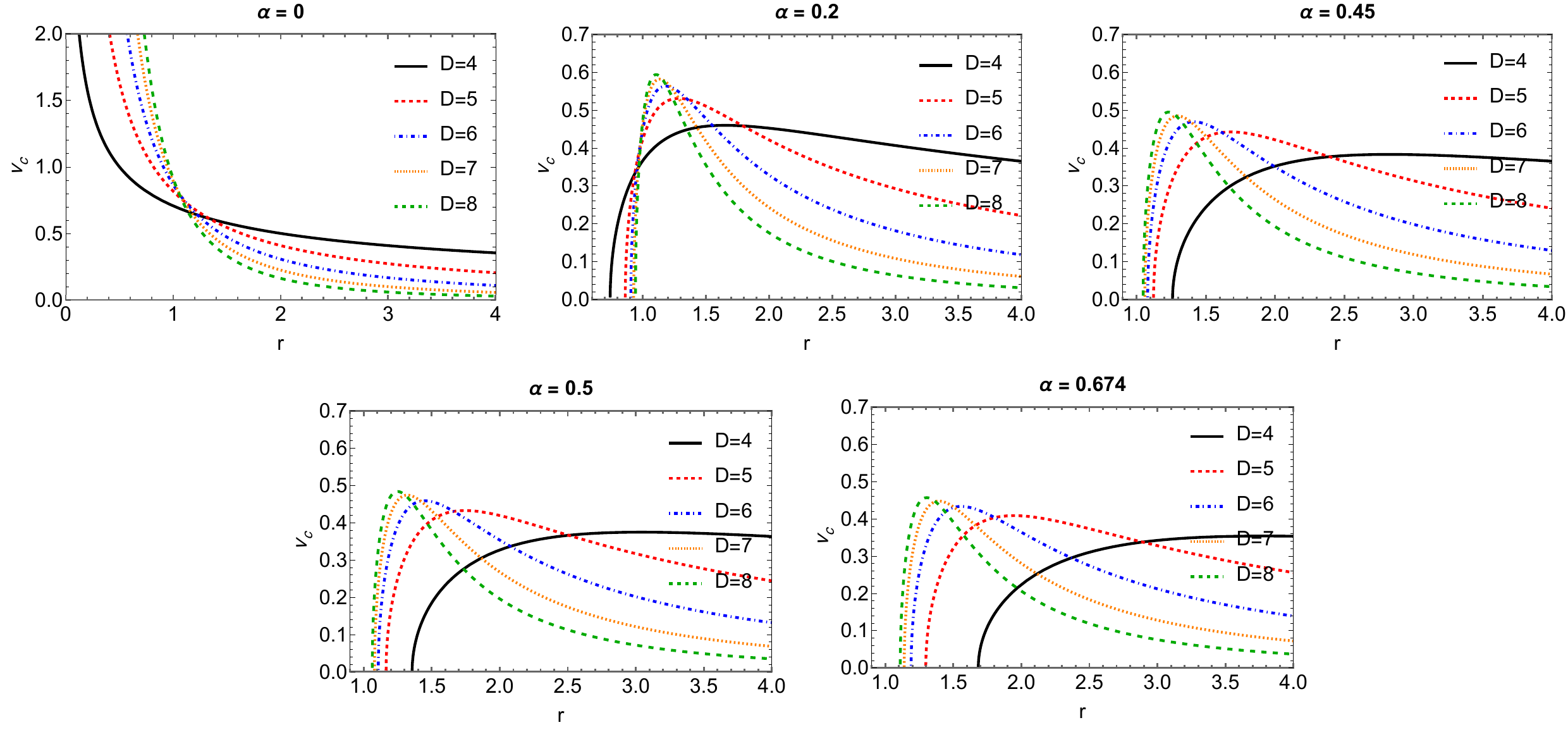}
\label{fig:vc-alpha}}
\caption{\small{\emph{The variation of $v_{c}$ with respect to $r$ for different values of $\alpha$ and spacetime dimensions. (a) illustrates the effect of changing $\alpha$ at fixed spacetime dimension, while panel (b) shows the impact of varying the spacetime dimension for constant $\alpha$.}}}
\label{fig:vc}
\end{figure}

The inflow's transition from subsonic to supersonic is indicated by the critical velocity $v_{c}(r)$. Its radial profile captures how the accretion process is controlled by the interaction of fluid dynamics, geometry, and gravity. According to the plots in Figure \ref{fig:vc}, increasing $\alpha$ makes gravity stronger, which raises the critical speed and moves the sonic point closer to the object. Higher spacetime dimensionality, on the other hand, decreases $v_{c}$ because the acceleration of the inflow is weakened by the dilution of gravitational strength in additional dimensions. Far from the compact object, the critical velocity naturally approaches zero in all cases. The behavior of $v_{c}(r)$ is especially important because it controls the accretion rate and sets the conditions for shock formation and energy release in the disk. Therefore, Figure \ref{fig:vc} demonstrates that both MOG corrections and extra-dimensional effects leave measurable imprints on the transonic structure of the flow, which could, in principle, distinguish higher-dimensional MOG compact objects from their general relativistic counterparts.

The adiabatic sound speed is defined as
\begin{equation}\label{a2}
a^2=\frac{dp}{d\rho}=w.
\end{equation}

Using Eqs. \eqref{v} and \eqref{a2}, one can derive the following equation for the adiabatic sound speed
\begin{equation}\label{lcs}
a^{2}=\frac{C_2}{\sqrt{f\left(r\right)+v^2}}-1\,,
\end{equation}
where $C_{2}$ is a constant defined in Eq. \eqref{C2}.

\section{Summary and conclusions}\label{Conclu}

In this work, we have explored the dynamics of neutral particles and the accretion process around a higher-dimensional, regular, spherically symmetric dark compact object within the Scalar-Tensor-Vector Gravity (MOG) framework. Our study addressed both the geodesic motion of test particles and the hydrodynamical accretion of a perfect fluid, with particular attention to the effects introduced by extra spatial dimensions.\\
We determined the effective potential, stable circular orbits, and the corresponding innermost stable circular orbit (ISCO) for the motion of the test particle. According to numerical analysis, matter can orbit closer to the compact object when the ISCO radius decreases as the number of spacetime dimensions increases. Consequently, the accretion disk's energy flux, effective temperature, and differential luminosity are all enhanced. These results demonstrate that the energetic output of accretion disks in MOG spacetimes is amplified by higher dimensions.\\
We also analyzed the accretion disk's thermal characteristics and contrasted the effective temperature profile with the most recent Sgr A* observational data from the Event Horizon Telescope (EHT). According to the analysis, the expected deviations from the Schwarzschild case caused by extra-dimensional corrections and MOG fall within the sensitivity range of the EHT measurements that are currently in use. In particular, the changes are constant for four-dimensional spacetimes for all the mentioned values of MOG parameter in table \ref{FTL}, and for the values of the MOG parameter in the range $\alpha\geq 1.5$, the effects are consistent with EHT data in for $D=5$. According to this comparison, such higher-dimensional and modified gravity signatures might be detected or constrained by upcoming high-precision VLBI observations.\\
In addition, we developed analytical expressions for the four-velocity and proper energy density of a perfect fluid undergoing spherical accretion onto the higher-dimensional MOG compact object. The analysis of mass evolution and critical accretion parameters demonstrated that the influence of extra dimensions persists not only at the level of particle dynamics but also in the hydrodynamical properties of inflowing matter. Our analysis of the fluid dynamics around higher-dimensional MOG dark compact objects shows that the radial velocity, energy density, and critical (sonic) velocity of the inflowing matter are strongly affected by both the MOG parameter $\alpha$ and the number of spacetime dimensions $D$. Increasing $\alpha$ enhances the effective gravitational interaction, leading to higher inflow velocities, outward shifts in density profiles, and larger critical speeds. In contrast, increasing $D$ dilutes the gravitational field, suppressing the radial and critical velocities while steepening the variation of energy density.These results highlight that extra dimensions and MOG corrections leave distinct imprints on the transonic structure and thermodynamic properties of accretion, which could in principle manifest in observable signatures of accretion disks.\\
Overall, our findings demonstrate that the observable imprints of accretion processes in MOG spacetimes are systematically strengthened by additional dimensions. The improvement of flux and temperature profiles and the decrease of the ISCO radius suggest that the electromagnetic spectrum of accretion disks may be affected in ways that can be measured. These signatures, lying at the threshold of detectability with present-day instruments, provide a promising avenue for testing the interplay of modified gravity and extra-dimensional physics. Future observational advances, particularly with the EHT and next-generation interferometric facilities, may therefore open the path to constraining or revealing such extensions of General Relativity.

\begin{acknowledgments}

The authors would like to thank John W. Moffat for their fruitful comments and discussions, which improved the
quality of the paper, considerably. Also, the Work of K. Nozari and S. Saghafi is supported financially by the INSF of Iran under the grant number $4038520$.

\end{acknowledgments}

\end{document}